\documentclass[twocolumn,onecolappendix]{aastex631}
\pdfoutput=1	
\usepackage{graphicx}	
\usepackage{amsmath}	
\usepackage{physics}
\usepackage{CJK}
\usepackage{mathtools}
\usepackage{xcolor}
\usepackage{bm}
\graphicspath{{./}{figures/}}
\usepackage[T1]{fontenc}
\definecolor{THUPurple}{RGB}{101,0,153}

\shorttitle{{\tt GRACE-DG}: a dust grain collisional evolution DG solver}
\shortauthors{Yang, Chen \& Bai}

\begin{document}
\begin{CJK*}{UTF8}{gbsn}
\title{{\tt GRACE-DG}: A Discontinuous Galerkin Method-Based Code for General Nonlinear Coagulation-Fragmentation Equations}


\author[0009-0008-1149-4217]{Jing Yang (杨晶)}
\affiliation{Institute for Advanced Study, Tsinghua University, Beijing 100084, China}
\email{jing-yan24@mails.tsinghua.edu.cn}

\author[0000-0001-7420-9606]{Zhuo Chen (陈卓)}
\affiliation{Institute for Advanced Study, Tsinghua University, Beijing 100084, China}
\email{chenzhuo\_astro@mail.tsinghua.edu.cn}

\author[0000-0001-6906-9549]{Xue-Ning Bai (白雪宁)}
\affiliation{Institute for Advanced Study, Tsinghua University, Beijing 100084, China}
\affiliation{Department of Astronomy, Tsinghua University, Beijing 100084, China}
\email{xbai@mail.tsinghua.edu.cn}

\begin{abstract}
    Dust plays a crucial role in protoplanetary disks (PPDs) evolution and planet formation, influencing disk dynamics through gas-dust coupling, regulating disk temperature by dominating continuum opacity, and altering disk ionization fraction by capturing free electrons. In this work, we develop a high-order discontinuous Galerkin (DG) method-based open-source code {\tt GRACE-DG} to solve the collision-induced coagulation-fragmentation equations.
    In particular, we have derived a new conservative formulation for the non-linear fragmentation term, which enables the DG method to capture the mass transfer process. The new solver exhibits good convergence in coupled aggregation and breakage simulations, making it highly suitable for future integration into hydrodynamic codes.
\end{abstract}

\keywords{Computational methods (1965) --- Dust physiscs (2229) --- Planet formation (1241)}

\section{Introduction}\label{sec:intro}
As the building blocks of planets, solid particles in protoplanetary disks (PPDs) must grow by more than ten orders of magnitude in size within a few million years.
According to the widely accepted core accretion model \citep{Bodenheimer1986,Pollack1996,Ida2004}, dust grains initially grow through coagulation into millimeter-sized pebbles.
However, this collisional growth is soon limited by the fragmentation and drift barriers.
To overcome these challenges, the streaming instability (SI) \citep{Youdin2005,Johansen2007} emerges as a key mechanism, concentrating solids and facilitating the gravitational collapse of dust clumps into planetesimals.
In particular, the threshold for dust clumping via SI depends sensitively on the dust size distribution \citep{Bai2010,LiRixin2021}, motivating a more detailed treatment beyond single-size approximation.
Therefore, grain growth in the early stages is crucial for subsequent planetesimal formation and also the pebble accretion \citep{Ormel2010}.

In addition to fueling planet formation, dust size distribution is a fundamental component in PPDs and plays a key role in a wide range of physical processes. Dust grains dominate the disk's opacity, which decreases significantly as grains grow, thereby affecting the disk's radiative transfer and overall thermodynamic structure \citep{Dullemond2004,Woitke2009}. Besides, small grains are efficient in capturing free electrons, altering the ionization state of the disk \citep{Okuzumi2009,Ivlev2016,Marchand2023}. The thermodynamic conditions and magnetic coupling in PPDs influence the local turbulence state, which in turn, regulate the dust size evolution \citep{Terquem2008,OH2012,Delage2022}. To capture the complex interplay of these processes, numerical simulations that couple dust-gas dynamics with realistic dust coagulation and fragmentation physics are essential for advancing our understanding of the early stages of planet formation \citep{Dominik2007,Birnstiel2010}.

However, three-dimensional hydrodynamic simulations with an evolving dust size distribution are numerically costly. \cite{Vericel2021} implemented dust growth and fragmentation in a 3D smoothed particle hydrodynamics (SPH) code, but their approach considered only a single dust species locally. In contrast, \cite{Stammler2022} modeled the 1D gas-dust evolution using a full local dust mass distribution, accounting for both collisional growth and radial transport. Additionally, attempts have been undertaken to simplify the coagulation process through the use of sub-grid models \citep{BKE2012,Tamfal2018,Vorobyov2018,Pfeil2024} and machine-learning-aided techniques \citep{Pfeil2022}. 
Despite these advances, a key challenge remains: developing an efficient method to solve the coagulation-fragmentation equations, thus enabling dust collision simulations on top of multidimensional hydrodynamic frameworks.

Pioneering efforts have been made in previous studies to apply the discontinuous Galerkin (DG) method to solve the Smoluchowski equations \citep{Liu2019,Lombart2021,Lombart2022,Lombart2024}. This high-order scheme has shown promising performance, efficiently solving the coagulation equation with a reduced number of mass bins compared to other discretized methods \citep{Nakagawa1981,Weidenschilling1980,Brauer2008,Birnstiel2010,Chaenoz2012}. Owing to its accuracy and computational efficiency, this methodology has recently enabled the first three-dimensional magnetohydrodynamic (MHD) simulation that self-consistently accounts for both dust dynamics and dust growth \citep{Lombart2026}. Despite these successes, the DG fragmentation solver developed by \cite{Lombart2024} restricts its applicability to pure destructive fragmentation. In this work, we introduce {\tt GRACE-DG} (GRAin Collisional Evolution - Discontinuous Galerkin)\footnote{\url{https://github.com/jing-xps/GRACE-DG}}\citep{gracedg2026}. By reformulating the fragmentation flux analytically, we extend the DG framework to incorporate both coagulation and general fragmentation processes, including diverse collisional fragmentation outcomes ranging from destructive fragmentation to mass transfer.

The structure of this paper is as follows. Section \ref{sec:eqs} presents the derivation of the conservative form of the Smoluchowski equations, which serve as the basis for the DG scheme. In Section \ref{sec:numerics}, we describe the numerical implementation in detail. The numerical results are shown in Section \ref{sec:results}. Finally, we provide further discussion in Section \ref{sec:discussion} and conclude in Section \ref{sec:conclusion}.

\section{Equations}\label{sec:eqs}
\begin{figure}
    \centering
    \includegraphics[width=1\linewidth]{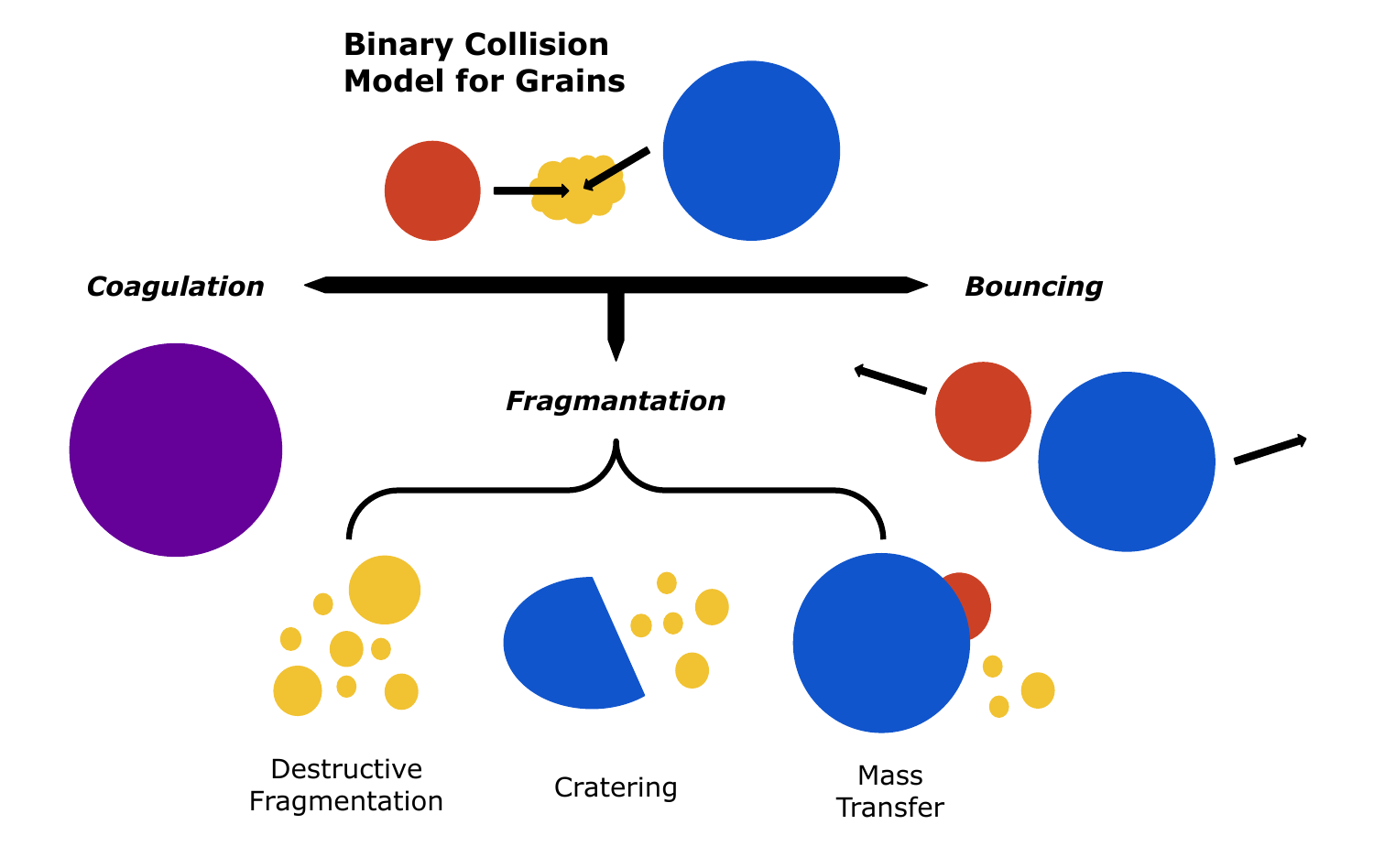}
    \caption{Illustration of collision outcomes. Depending on the relative velocity between colliding particles, three main outcomes are possible: coagulation, bouncing, and fragmentation. In the case of fragmentation, the specific outcome further depends on the mass ratio of the colliding grains. (for example, Equation \ref{eq:mfrag})}
    \label{fig:model}
\end{figure}

The equations considered in this study describe the time evolution of dust size distribution under the combined effects of binary aggregation and multiple breakage. As illustrated in Figure \ref{fig:model}, collisions between particles lead to different outcomes. Among these, bouncing does not alter the global size distribution. Cratering and mass transfer are types of partial fragmentation. In these events, a large remnant of the original dust particle (shown in blue) survives the collision. We present the governing equations for coagulation and fragmentation in Section \ref{sec:eqs/coag} and Section \ref{sec:eqs/frag}.

\subsection{Coagulation}\label{sec:eqs/coag}
Mass conservation for a distribution of growing grains can be formulated by the Smoluchowski equation \citep{1916ZPhy...17..557S},
\begin{equation}\label{eq:fcoag}
\begin{split}
     \frac{\partial f(x,\tau)}{\partial \tau}=&\frac{1}{2}\int_0^x\mathcal{K}_{\mathrm{coag}}(x-y,y)f(x-y,\tau)f(y,\tau)\,\mathrm{d}y\\
    &-f(x,\tau)\int_0^{\infty}\mathcal{K}_{\mathrm{coag}}(x,y)f(y,\tau)\,\mathrm{d}y,
\end{split}
\end{equation}
where $f(x,\tau)$ is the grain number density per unit mass of mass $x$ at time $\tau$, $\mathcal K_{\mathrm{coag}}(x,y)$ is the coagulation kernel, denoting a collision rate of coagulation 
between two grains of mass $x$ and mass $y$. Thus it has the units of $[\rm{length}^3/\rm{time}]$. The first term on the right-hand-side (RHS) of Equation \ref{eq:fcoag} represents the production of particles with mass $x$ through binary collisions, while the second term accounts for the loss of $x$-mass particles due to collisions with other grains.

The Smoluchowski equation can be rewritten in to a conservative form using the mass density $g(x,\tau) = xf(x,\tau)$ \citep{Tanaka1996},
\begin{equation}\label{eq:gcoag}
    \frac{\partial g(x,\tau)}{\partial \tau}+\frac{\partial F_{\mathrm{coag}}[g](x,\tau)}{\partial x}=0,
\end{equation}
\begin{equation}\label{eq:Fcoag}
    F_{\mathrm{coag}}[g](x,\tau) = \int_0^x\mathrm{d}u\int_{x-u}^{\infty}\mathrm{d}v\, \mathcal{K}_{\rm{coag}}(u,v)g(u,\tau)\frac{g(v,\tau)}{v},
\end{equation}
where $F_{\mathrm{coag}}[g](x,\tau)$ denotes the flux of $g(x,\tau)$ across mass coordinate $x$ induced by coagulation. Some intuition can be gained by considering the flux across $x$ as arising from collisions between particles of mass $u<x$ and particles of mass $v$, such that the resulting mass $u+v>x$.

Unlike conventional fluxes in hydrodynamics, which typically only depend on neighboring cells, $F_{\mathrm{coag}}[g]$ is nonlocal---it depends on the global distribution of $g(x,\tau)$, which is highlighted by $[g]$. To solve this scalar hyperbolic conservation equation, we employ a DG method tailored to handle the nonlocal nature of the flux.

\subsection{Fragmentation}\label{sec:eqs/frag}
In addition to sticking upon collision, two colliding particles may also fragment into smaller grains when the collision velocity exceeds a certain threshold. The mass distribution of grains where mass transfers from large to small grains through this fragmentation process is modeled by the fragmentation equation. \cite{Liu2019} first considered a linear fragmentation model,
\begin{equation}\label{eq:f19}
    \frac{\partial f(x,\tau)}{\partial \tau}=\int_x^{\infty}b(x,y)S(y)f(y,\tau)\, \mathrm{d}y-S(x)f(x,\tau),
\end{equation}
where $S(x)$ describes the rate at which particles of mass $x$ are selected to break, and the breakage function $b(x,y)$ is the probability density function for the formation of particles of mass $x$ from a parent particle of mass $y$. In this linear model, fragmentation is treated as a single-particle process, independent of any collisional partner. The integration domain $y \ge x$ reflects the physical constraint that only particles with mass larger than $x$ can produce fragments of mass $x$.

\cite{Lombart2022} further considered a non-linear collision-induced fragmentation model,
\begin{equation}\label{eq:f22}
\begin{split}
    \frac{\partial f(x,\tau)}{\partial \tau} = &\int_0^{\infty} \mathrm{d}z
    \int_x^{\infty}\mathrm{d}y\, \mathcal{K}_{\mathrm{frag}}(y,z)\tilde b(x,y;z)f(y,\tau)f(z,\tau)\\
    &-f(x,\tau)\int_0^{\infty}\mathcal K_{\mathrm{frag}}(x,y)f(y,\tau)\, \mathrm{d}y,
\end{split}
\end{equation}
where $\mathcal{K}_{\mathrm{frag}}(x,y)$ is the fragmentation kernel and $\tilde b(x,y;z)$ is the breakage kernel describing the distribution of fragments of mass $x$ produced from a particle of mass $y$ in a collision with a particle of mass $z$. It is important to note that Equation \ref{eq:f22} relies on the underlying assumption that, during a collision between particles of masses $y$ and $z$, only particle $y$ undergoes fragmentation while particle $z$ remains intact. Consequently, the breakage kernel is defined on $x\in (0,y]$ and satisfies the local mass conservation condition $\int_0^{y} x\, \tilde b(x,y;z)\,\mathrm{d}x = y$. While this assumption simplifies the modeling, it restricts the physical realism of the formulation, as it neglects the possibility that both colliding particles may fragment.

A more general non-linear fragmentation equation was later considered by \cite{Lombart2024},
\begin{equation}\label{eq:f24}
\begin{split}
    \frac{\partial f(x,\tau)}{\partial\tau} = &\frac{1}{2}\int_0^{\infty}\mathrm{d}z\int_0^{\infty}\mathrm{d}y\,\mathtt{1}_{y+z\ge x}b(x,y,z)\mathcal K_{\mathrm{frag}}(y,z) \\
    &\quad\quad\quad\quad\quad\times  f(y,\tau)f(z,\tau)\\
    &-f(x,\tau)\int_0^{\infty}\mathcal K_{\mathrm{frag}}(x,y)f(y,\tau)\, \mathrm{d}y,
\end{split}
\end{equation}
which accounts for a more realistic collision outcome by allowing the total mass of the two colliding particles to be redistributed among fragments. In this formulation, the breakage kernel $b(x,y,z)$ is defined on $x\in(0,y+z]$ and satisfies $\int_0^{y+z} x\, b(x,y,z)\,\mathrm{d}x = y+z$. The indicator function $\mathtt{1}_{y+z \ge x}$ equals 1 when $y+z \ge x$ and 0 otherwise, ensuring that fragments of mass $x$ can only be produced from collisions whose combined mass is sufficient. Similar to Equation \ref{eq:fcoag}, the first term on the RHS represents the formation of particles of mass $x$, while the second term describes the loss of particles of mass $x$ due to collisions with other grains, occurring at a rate governed by the fragmentation kernel.

To solve Equation \ref{eq:f24} within the DG scheme, \cite{Lombart2024} derived the conservative form of this general non-linear fragmentation equation:
\begin{equation}\label{eq:g24}
    \frac{\partial g(x,\tau)}{\partial \tau}+\frac{\partial F_{\mathrm{frag}}[g](x,\tau)}{\partial x}=0,
\end{equation}
with the associated fragmentation flux
\begin{equation}\label{eq:F24}
\begin{split}
    F_{\mathrm{frag}}[g](x,\tau) 
    = & -\frac{1}{2} \int_0^x\mathrm{d}x'\int_0^{\infty}\mathrm{d}z\int_0^{\infty}\mathrm{d}y\, \mathtt{1}_{y+z\ge x} \frac{x'}{yz} \\
     &\times \mathcal{K}_{\mathrm{frag}}(y,z)b(x',y,z)g(y,\tau) g(z,\tau)  \\ 
     & +  \int_0^x\mathrm{d}z\int_{0}^{\infty}\mathrm{d}y \,  \mathtt{1}_{y+z\ge x}\frac{1}{y}  \\
     &\times\mathcal{K}_{\mathrm{frag}}(y,z) g(z,\tau) g(y,\tau).
\end{split}
\end{equation}
As reported in \cite{Lombart2024}, the numerical evaluation of $F_{\mathrm{frag}}[g]$ is performed using Gaussian quadrature. This numerical integration fails when the breakage kernel $b(x,y,z)$ incorporates a Dirac delta function, which is required to model cratering or mass transfer processes where $b(x,y,z)=A x^{\alpha} + \delta(x-m_{\mathrm{left}})$\citep{Kobayashi2010,Hirashita2021}.
(See Equation \ref{eq:break} in Section \ref{sec:coupled} for a more detailed explanation of this breakage kernel). To bypass the delta function, the integration in Equation \ref{eq:F24} must be reordered to $\int\mathrm{d}z\int\mathrm{d}y\int\mathrm{d}x'$, allowing the innermost integral over the fragment mass $x'$ to be evaluated analytically beforehand.

In order to overcome this limitation, we introduce a novel conservative formulation of the mass flux $F_{\mathrm{frag}}[g](x,\tau)$ that consistently accommodates general fragmentation processes; a rigorous derivation is presented in Appendix \ref{sec:proof}.


\begin{equation}\label{eq:Ffrag}
\begin{split}
    F_{\mathrm{frag}}[g](x,\tau)=&
    \underbrace{
        \begin{aligned}
            \int_0^x\mathrm{d}u\int_{x-u}^{\infty}\mathrm{d}v\int_x^{u+v}\, \frac{w\mathrm{d}w}{v(u+v)} \\
            \times \mathcal K_{\mathrm{frag}}(u,v)b(w,u,v)g(u,\tau)g(v,\tau)
        \end{aligned}
    }_{\text{positive flux}}\\
    &
    \underbrace{
        \begin{aligned}
            -\int_x^{\infty}\mathrm{d}u\int_0^{\infty}\mathrm{d}v\int_0^x\, \frac{w\mathrm{d}w}{v(u+v)} \\
            \times\mathcal K_{\mathrm{frag}}(u,v)b(w,u,v) g(u,\tau)g(v,\tau)
        \end{aligned}
    }_{\text{negative flux}}.
\end{split}
\end{equation}

\begin{figure}
\centering
    \includegraphics[width=1\linewidth]{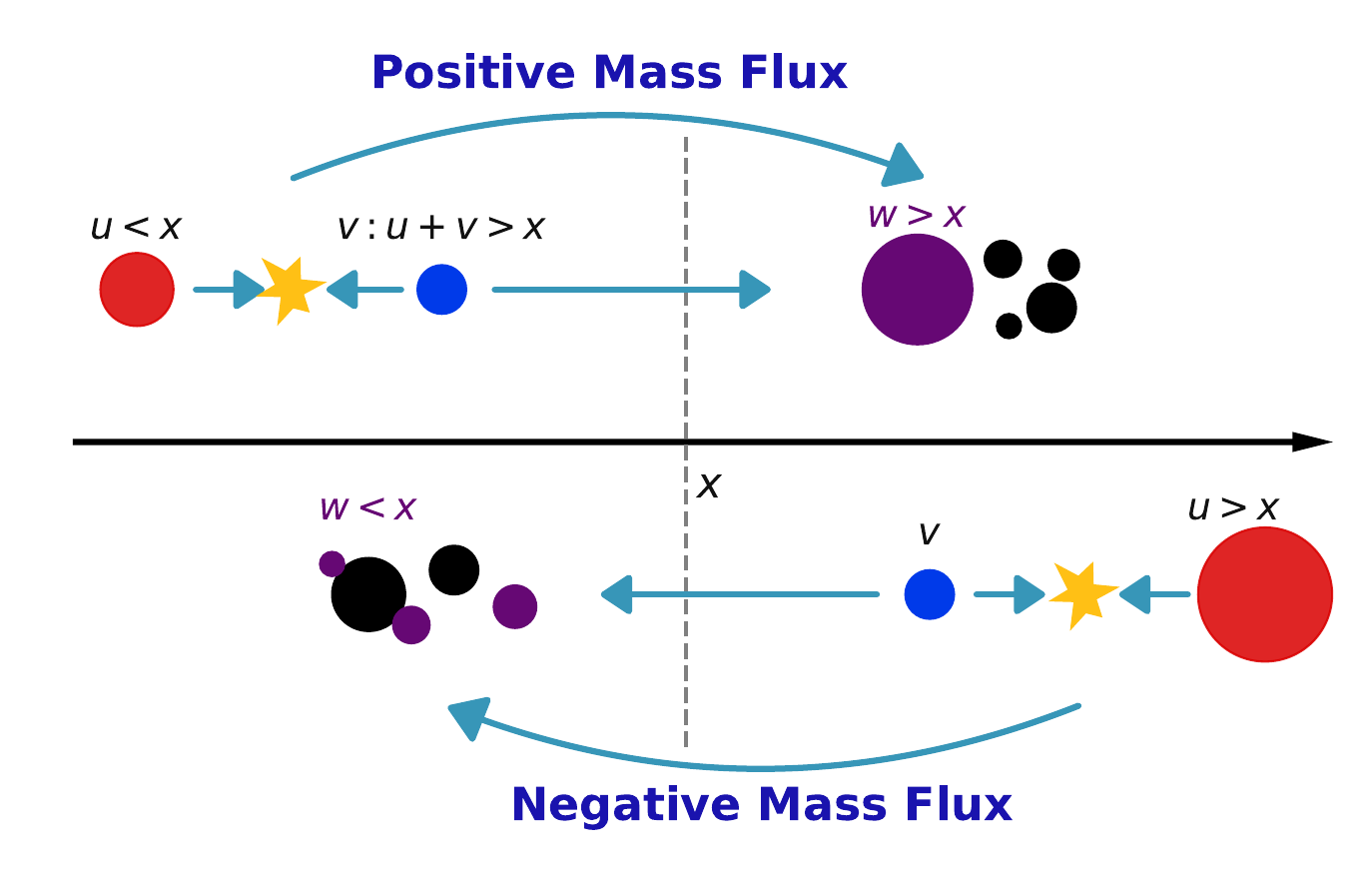}
    \caption{Schematic illustration on the physical interpretation of the fragmentation flux in Equation \ref{eq:Ffrag}. The upper panel shows the processes inducing the positive flux, while the lower panel depicts the events responsible for the negative flux.}
    \label{fig:fragflux}
\end{figure}

In this formulation, the innermost integration is performed with respect to the fragment mass $w$, and all physical constraints are explicitly encoded in the integration limits. As illustrated in Figure \ref{fig:fragflux}, some physical intuition can also be gained by considering the origin of the flux: a positive flux is generated when a particle of mass $u<x$ collides with a particle of mass $v$ such that the total mass $u+v>x$, resulting in the formation of a particle with mass $w>x$. Conversely, a negative flux is associated with collisions in which a particle of mass $u>x$ is involved, producing a fragment with mass $w<x$. Noticeably, we could consider the coagulation process as a special case of this general fragmentation formalism by forbidding the collided particles to fragment, which is equivalent to setting $b(w,u,v) = \delta\left(w-(u+v)\right)$.

\section{Numerics}\label{sec:numerics}
We present the detailed implementation of the numerical algorithm used to solve the coagulation-fragmentation equations.
\subsection{Discontinuous Galerkin Method}
The discontinuous Galerkin (DG) method is a finite element approach that employs completely discontinuous basis functions, typically chosen as piecewise polynomials \citep{cockburn_rungekutta_2001}. DG methods have been successfully applied to first-order hyperbolic problems, particularly for convection-dominated flows and local hyperbolic conservation laws \citep{cockburn_rungekutta_2001}. In this work, we apply it to solve the following integro-differential equation:
\begin{equation}\label{eq:2.1}
    \frac{\partial g(x,\tau)}{\partial \tau}+\frac{\partial F[g](x,\tau)}{\partial x}=0.
\end{equation}

First, let us decompose the domain of interest [$x_{\mathrm{min}}, x_{\mathrm{max}}$] into $N$ subintervals. Each mass bin is defined by $I_j = [x_{j-1/2},x_{j+1/2}],\ j\in[0,N-1]$. The size of $j-$th cell is defined as $h_j = x_{j+1/2}-x_{j-1/2}$, and it's centered around the position $x_j = (x_{j+1/2}+x_{j-1/2})/2$. Since the difference between $x_{\max}$ and $x_{\min}$ typically spans several orders of magnitude, we adopt a logarithmic grid to efficiently resolve the wide range of particle masses.

Next, we obtain the weak formulation of Equation \ref{eq:2.1} by multiplying by a test function $\phi(x)$, integrating over $I_j$ for each subinterval:
\begin{equation}\label{eq:2.2}
    \begin{split}
        \int_{I_j}\frac{\partial g_j}{\partial \tau}\phi\, \mathrm{d}x &-\int_{I_j}F[g](x,\tau)\frac{\partial \phi}{\partial x}\, \mathrm{d}x\\
        &+F[g](x_{j+1/2},\tau)\phi(x_{j+1/2})\\
        &-F[g](x_{j-1/2},\tau)\phi(x_{j-1/2}) =0.
    \end{split}
\end{equation}
Then, within each cell, we approximate the solution $g(x,t)$ by a linear combination of Legendre polynomials of degree at most $k$:
\begin{equation}\label{eq:2.3}
    \forall x\in I_j \quad g(x,\tau)\approx g_j(x,\tau) = \sum_{i=0}^k g_j^i(\tau)\phi_i(\xi_j(x)),
\end{equation}
where $g_j^i(\tau)$ is the component of $g_j(x,\tau)$ on the $i-$Legendre basis:
\begin{equation}\label{eq:2.4}
    \phi_0(\xi) = 1,\quad \phi_1(\xi) = \xi,\quad \phi_2(\xi) = \frac{1}{2}(3\xi^2-1),\quad....
\end{equation}
And $\xi_j(x) = \frac{2}{h_j}(x-x_j)$ is used to map $x\in I_j$ onto $\xi_j \in [-1,1]$. In this way, the mass $x$ and time $\tau$ variables are decoupled. 

In addition, the orthogonality of the Legendre functions gives,
\begin{equation}\label{eq:2.5}
    \int_{-1}^1\phi_i(\xi)\phi_k(\xi)\, \mathrm{d}\xi= \delta_{ik}\cdot d_i, \quad d_i = \frac{2}{2i+1}.
\end{equation}
Combining Equation \ref{eq:2.2}, Equation \ref{eq:2.3} and Equation \ref{eq:2.5}, and letting the test function in Equation \ref{eq:2.2} to be Legendre polynomials, one obtains equations with respect to the coefficients $\mathbf{g}_j=(g_j^0, g_j^1,...,g_j^k)^T$. We write it in a compact form,
\begin{equation}\label{eq:bfg}
    \frac{d\mathbf{g}_j(\tau)}{d\tau}=\mathbf{L}[g],
\end{equation}
where $\mathbf{L}$ is an operator acting on the global mass density distribution $g(x, \tau)$,
\begin{equation}\label{eq:evo}
    \begin{split}
    \mathbf{L}[g] = \frac{2}{h_j}\cdot &
    \begin{bmatrix}
    \scriptstyle 1/d_0 & &  \\
    & \ddots & \\
     & & \scriptstyle 1/d_k
    \end{bmatrix} \\
    & \cdot \left( \begin{aligned}
        & \int_{I_j}F[g](x,\tau)\partial_x \bm{\phi} (\xi_j(x))\mathop{}\!\mathrm{d}x \\
        & - F[g](x_{j+1/2},\tau)\bm{\phi}(\xi_j(x_{j+1/2})) \\
        & + F[g](x_{j-1/2},\tau)\bm{\phi}(\xi_j(x_{j-1/2}))
    \end{aligned} \right),
    \end{split}
\end{equation}
where $\bm{\phi}=(\phi_0, \phi_1,...,\phi_k)^T$ denotes the vector of basis functions. As a result, the original system of partial differential equations (PDEs) is reduced to a system of ordinary differential equations (ODEs) onto the coefficients $g_j^i(\tau)$.

\subsection{Algorithm Flowchart}
We implement the following flowchart to solve the non-linear coagulation-fragmentation equation, with the flux term evaluated by Gaussian quadrature. The positivity of cell-averaged mass density is guaranteed by the CFL condition, while a scaling limiter and reconstruction help to preserve the positivity of the global solution \citep{Liu2019}. For the higher-order time discretization, we employ the strong stability preserving Runge-Kutta (SSPRK) third-order method \citep{Gottlieb21}.

\subsubsection{Initialization}
From the initial mass density distribution $g_0(x)=g(x,0)$, we generate $\mathbf{g}_j$ at $\tau=0$ by
$$
\forall\ j\in [0, N-1],\quad \forall\ i \in [0,k]
$$
\begin{equation}\label{eq:init}
    g_j^i(0)=\frac{1}{d_i}\int_{-1}^1g_0(\frac{h_j}{2}\xi_j+x_j)\phi_{i}(\xi_j)\, \mathrm{d}\xi_j.
\end{equation}

\subsubsection{Reconstruction}
The physical meaning of mass density $g(x,\tau)$ requires $g_j(x,\tau)\ge 0,\ \forall\ x\in I_j$. This is achieved by a linear scaling reconstruction in each cell, using cell average as a reference \citep{Liu2019,Lombart2022}:
\begin{equation}\label{eq:recon}
    \tilde g_j(x) = \gamma_j (g_j(x)-\bar g_j) + \bar g_j,
\end{equation}
where the scaling limiter $\gamma_j$ is defined by
\begin{equation}\label{eq:gamma}
    \gamma_j = \min\left\{ 1,\frac{\bar g_j}{\bar g_j - \underset{x \in I_j}{\min} g_j(x)} \right\}.
\end{equation}
where $\bar{g}_j$ denotes the cell average of $g_j(x)$ over the interval $I_j$. It can be readily shown that $\bar{g}_j$ corresponds to the zeroth coefficient $g_j^0$ in the polynomial expansion,
\begin{equation}\label{eq:average}
    \begin{split}
        \bar g_j &=\frac{1}{h_j}\int_{I_j}g_j(x)\,\mathrm{d}x=\frac{1}{h_j}\int_{I_j}\sum_{i=0}^kg_j^i\cdot \phi_i(\xi_j(x))\,\mathrm{d}x\\  
        &=\frac{1}{h_j}\sum_{i=0}^kg_j^i\cdot \int_{-1}^1\phi_i(\xi_j)\cdot\frac{h_j}{2}\,\mathrm{d}\xi_j=g_j^0.
    \end{split}
\end{equation}
The minimum mass density $\underset{x \in I_j}{\min} g_j(x)$ is calculated analytically. It's easy to check that the cell average of the reconstructed function $\tilde g_j(x)$ is still $g_j^0$. The resulting $\tilde g_j(x)$ therefore constitutes a scaled polynomial approximation and is used as the solution representation in the subsequent evolution step.

\subsubsection{Evolution}
According to the ODE system, Equation \ref{eq:evo}, forward Euler discretization yields the update rule for the zeroth coefficients (cell averages $\bar g_j$) as,
\begin{equation}\label{eq:g0}
    (g_j^0)^{n+1} = (g_j^0)^n +\frac{\Delta t}{h_j}\cdot [F[g](x_{j-1/2})- F[g](x_{j+1/2})],
\end{equation}
where $h_j=x_{j+1/2}-x_{j-1/2}$ is the size of $j$-th cell. The Courant-Friedrichs-Lewy condition (CFL) of the scheme is chosen to guarantee the positivity of the cell averages $(g_j^0)^{n+1}>0$ \citep{Liu2019,Lombart2021}:
\begin{equation}\label{eq:dt}
    \Delta t< \frac{h_j\cdot(g_j^0)^n}{|F[g](x_{j-1/2})- F[g](x_{j+1/2})|}.
\end{equation}
In addition, we use multi-step time discretization to achieve higher-order accuracy in time while maintaining the strong stability property. An optimal third-order strong stability preserving Runge-Kutta (SSPRK) method with a CFL coefficient $c=1$ is given by \citep{Gottlieb21,Lombart2021},
\begin{eqnarray}\label{eq:RK3}
    \mathbf{g}_j^{(1)} &=& \mathbf{g}_j^n +\Delta t\mathbf{L}[g^n], \nonumber \\
    \mathbf{g}_j^{(2)} &=& \frac{3}{4}\mathbf{g}_j^n +\frac{1}{4}\left(\mathbf{g}_j^{(1)}+\Delta t\mathbf{L}[g^{(1)}]\right), \\
    \mathbf{g}_j^{n+1} &=& \frac{1}{3}\mathbf{g}_j^n +\frac{2}{3}\left(\mathbf{g}_j^{(2)}+\Delta t\mathbf{L}[g^{(2)}]\right). \nonumber
\end{eqnarray}

We summarize the algorithm flowchart into the following steps:
\begin{enumerate}
    \item Initialization: generate $\forall\ j\in [0, N-1]$, $\forall\ i\in[0,1,...,k]$, $g_j^i(0)$ from the initial mass density distribution $g_0(x)$ by Equation \ref{eq:init}.
    \item Reconstruction: use Equation \ref{eq:recon} to reconstruct $g_j(x)$, and continue with step \ref{3}. \label{2}
    \item Evolution: use Equation \ref{eq:RK3} to compute $\forall\ j\in [0, N-1]$, $\forall\ i\in[0,1,...,k]$, $(g_j^i)^{n+1}$, and continue with step \ref{2}. \label{3}
\end{enumerate}

\subsection{Flux Evaluation}\label{sec:flux}
The implementation of operator $\mathbf{L}[g]$ in Equation \ref{eq:evo} involves two key components: i) the value of the flux at each cell interface; ii) the integral of the flux over each cell. Considering a mass interval $[x_{\mathrm{min}}, x_{\mathrm{max}}]$, we truncate the mass density fluxes in Equations \ref{eq:Fcoag} and \ref{eq:Ffrag} as:
\begin{equation}\label{eq:Fcoagtrun}
\begin{split}
    F_{\mathrm{coag}}^{\rm{tr}}[g](x,\tau) = \int_{x_{\mathrm{min}}}^x\mathrm{d}u\int_{x-u+x_{\mathrm{min}}}^{x_{\mathrm{max}}+\Delta x} \mathrm{d}v\\
    \times \mathcal{K}_{\mathrm{coag}}(u,v) g(u,\tau)\frac{g(v,\tau)}{v},
\end{split}
\end{equation}

\begin{equation}\label{eq:Ffragtrun}
\begin{split}
    F_{\mathrm{frag}}^{\rm{tr}}[g](x,\tau)&=\int_{x_{\mathrm{min}}}^x\mathrm{d}u\int_{x-u+x_{\mathrm{min}}}^{x_{\mathrm{max}}+\Delta x}\mathrm{d}v\int_{x}^{u+v} \frac{w\mathrm{d}w}{v(u+v)} \\
    &\times b(w,u,v)\mathcal K_{\mathrm{frag}}(u,v)g(u,\tau)g(v,\tau)\\
    &-\int_x^{x_{\mathrm{max}}}\mathrm{d}u\int_{x_{\mathrm{min}}}^{x_{\mathrm{max}}+\Delta x}\mathrm{d}v\int_{x_{\mathrm{min}}}^x \frac{w\mathrm{d}w}{v(u+v)} \\
    &\times b(w,u,v)\mathcal K_{\mathrm{frag}}(u,v)g(u,\tau)g(v,\tau).
\end{split}
\end{equation}

Here, $\Delta x$ serves as an adjustment parameter for the upper integration limit of the colliding particle mass $v$, dictating how interactions at the upper boundary of the mass domain are handled. The choice of $\Delta x$ has an impact on the physical property of the fluxes. When $\Delta x=-u+x_{\rm{min}}$, Equation \ref{eq:Fcoagtrun} and \ref{eq:Ffragtrun} ensures that $F^{\rm{tr}}_{\rm{coag}}(x_{\mathrm{min}})=F^{\rm{tr}}_{\rm{frag}}(x_{\mathrm{min}})=F^{\rm{tr}}_{\rm{coag}}(x_{\mathrm{max}})=F^{\rm{tr}}_{\rm{frag}}(x_{\mathrm{max}})=0$, and the mass in $[x_{\mathrm{min}}, x_{\mathrm{max}}]$ will be conserved. We call them ``conservative fluxes''. When $\Delta x=0$, the integration is performed on the whole computational domain $[x_{\mathrm{min}}, x_{\mathrm{max}}]$ uniformly, and we call them ``non-conservative fluxes''. Consequently, for hydrodynamic simulations that require mass conservation, setting $\Delta x=-u+x_{\rm{min}}$ is more appropriate; for comparison with analytic solutions, which assume $x\in(0,\infty)$, setting $\Delta x=0$ may be a more appropriate choice \citep{Lombart2021}. In this work, $\Delta x =0$ is our default choice unless explicitly stated.


For the numerical integration of the fluxes, we use Gaussian quadrature of order $Q$ with the Gauss evaluation points $s_{\alpha}\in (-1,1)$ and the weights $w_{\alpha}>0$ \citep{Liu2019,Lombart2024}:
\begin{equation}\label{eq:quad}
\begin{split}
    \int_a^b f(x)\, \mathrm{d}x =&\frac{b-a}{2}\int_{-1}^1f(\frac{b+a}{2}+\frac{b-a}{2}s)\,\mathrm{d}s\\
    =&\frac{b-a}{2}\sum_{\alpha=1}^Qw_{\alpha}f(\frac{b+a}{2}+\frac{b-a}{2}s_{\alpha})\\
    &+\mathcal{O}\left((b-a)^{2Q}\right),
\end{split}
\end{equation}
where the approximation residual is zero when $f(x)$ is a polynomial of degree at most $2Q-1$. Our numerical experiments reveal optimal convergence at $Q=k+1$, rather than the expected $Q=k$, an observation that was also previously noted by \cite{Liu2019}. We therefore choose a quadrature number of $Q=k+1$ for all simulations. Taking the conservative coagulation flux ($F_{\mathrm{coag}}^{\mathrm{tr}}[g](x_{j+1/2},\tau)$ with $\Delta x=-u+x_{\rm{min}}$) as an example,
\begin{equation}
\begin{split}
    F_{\mathrm{coag}}^{\mathrm{tr}}[g](x_{j+1/2},\tau)&=\int_{x_{\mathrm{min}}}^{x_{j+1/2}}\mathrm{d}u\int_{x_{j+1/2}-u+x_{\mathrm{min}}}^{x_{\mathrm{max}}-u+x_{\rm{min}}}\mathrm{d}v\\
    &\times \mathcal{K}_{\mathrm{coag}}(u,v)g(u,\tau)\frac{g(v,\tau)}{v}\\
    &=\sum_{l=0}^{j}\frac{h_l}{2}\sum_{\alpha=1}^Qw_{\alpha}g(x_l^\alpha,\tau)\Gamma_l^{\alpha},
\end{split}
\end{equation}
where $h_l=x_{l+1/2}-x_{l-1/2}$ is the size of $l$-th cell, and $x_l^\alpha=x_l+h_l/2\cdot s_{\alpha}$. 
\begin{equation}
\begin{split}
    \Gamma_l^\alpha&=\int_{x_{j+1/2}-x_l^\alpha+x_{\mathrm{min}}}^{x_{\mathrm{max}}-x_l^\alpha+x_{\mathrm{min}}}\mathcal{K}_{\mathrm{coag}}(x_l^\alpha,v)\frac{g(v,\tau)}{v}\, \mathrm{d}v\\
    &= \sum_{m=A}^B\frac{b_m-a_m}{2}\sum_{\beta=1}^Qw_{\beta}\mathcal{K}_{\mathrm{coag}}(x_l^\alpha,x_m^\beta)\frac{g(x_m^\beta,\tau)}{x_m^\beta},
\end{split}
\end{equation}
where the index $A$ and $B$ are chosen such that $x_{j+1/2}-x_l^\alpha+x_{\mathrm{min}}\in I_A$, $x_{\mathrm{max}}-x_l^\alpha+x_{\mathrm{min}}\in I_B$. When $m=A$, $a_m = x_{j+1/2}-x_l^\alpha+x_{\mathrm{min}}$ and when $m=B$, $b_m=x_{\mathrm{max}}-x_l^\alpha+x_{\mathrm{min}}$. Otherwise, $a_m=x_{m-1/2}$, $b_m=x_{m+1/2}$. And the quadrature points are given by,
\begin{equation}
    x_m^\beta = \frac{b_m+a_m}{2}+\frac{b_m-a_m}{2}\cdot s_{\beta}.
\end{equation}

Now, let's consider the integral of the flux $F_{\mathrm{coag}}^{\mathrm{tr}}[g](x,\tau)$ over $I_j$,
\begin{equation}
\begin{split}
    \int_{I_j}F_{\mathrm{coag}}^{\mathrm{tr}}[g](x,\tau)\partial_x\phi \,\mathrm{d}x &=\int_{-1}^1F_{\mathrm{coag}}^{\mathrm{tr}}[g](\xi,\tau)\partial_\xi\phi \,\mathrm{d}\xi\\
    &=\sum_{\theta=1}^Qw_{\theta}\phi'(s_{\theta})F_{\mathrm{coag}}^{\mathrm{tr}}[g](x_j^\theta,\tau),
\end{split}
\end{equation}
where $x_j^\theta=x_j+h_j/2\cdot s_{\theta}$. The evaluation of $F_{\mathrm{coag}}^{\mathrm{tr}}[g](x_j^\theta,\tau)$ follows the same procedure as for $F_{\mathrm{coag}}^{\mathrm{tr}}[g](x_{j+1/2},\tau)$, with $x_{j+1/2}$ simply replaced by $x_j^\theta$.

An additional consideration arises when the breakage kernel $b(x,y,z)$ contains Dirac delta functions in $x$. The presence of these delta functions necessitates the analytical evaluation of the innermost integral in Equation \ref{eq:Ffragtrun}, which requires a detailed piecewise discussion depending on the value of $x$ (see Equation \ref{eq:inte1} and \ref{eq:inte2}). Consequently, the fragmentation flux $F_{\mathrm{frag}}[g](x,\tau)$ becomes discontinuous in $x$, which introduces challenges in evaluating $\int_{I_j}F_{\mathrm{frag}}^{\mathrm{tr}}[g](x,\tau) \partial_x\phi \, \mathrm{d}x$. This issue is circumvented by reordering the integration sequence, as detailed in Appendix \ref{sec:gf}.

\section{Results}\label{sec:results}
This section presents three numerical benchmark tests to assess the performance of the proposed DG scheme by comparing our numerical results with analytical solutions. The two pure coagulation tests, which utilize a constant kernel and an additive kernel, follow the setups of \cite{Liu2019} and \cite{Lombart2021}. For the additive kernel case, we compare our numerical results with the {\tt DustPy} code to demonstrate the DG scheme's ability to mitigate the over-diffusion problem often encountered in standard numerical methods. The pure fragmentation test follows the configuration detailed by \cite{Lombart2022}. In addition, we perform simulations that incorporate both aggregation and breakage processes to illustrate the convergence properties of the DG scheme.

\subsection{Diagnostics}\label{sec:error}
Numerical errors are measured using continuous and discrete norms, which are defined as \citep{Liu2019},
\begin{equation}
    e_{c,N}(\tau) = \sum_{j=0}^{N-1}\frac{h_j}{2}\sum_{\alpha=1}^{R}\omega_{\alpha}|g_j(x_j^{\alpha},\tau)-g(x_j^{\alpha},\tau)|,
\end{equation}
\begin{equation}
    e_{d,N}(\tau) = \sum_{j=0}^{N-1}h_j|g_j(\hat x_j,\tau)-g(\hat x_j,\tau)|,
\end{equation}
where $g_j(x)$ and $g(x)$ are the numerical and analytic solutions. $R=16$ is the number of Gauss points we take within $I_j$, $\omega_{\alpha}$ is the corresponding weights. $x_j^{\alpha} = x_j + h_j/2 \cdot s_{\alpha}$ and $\hat x_j = \sqrt{x_{j-1/2}x_{j+1/2}}$ is the geometric mean of the bin $I_j$. Formally, the errors are normalized by the integral of the analytical solution, yielding relative errors. However, for the benchmark tests considered here, the total mass of the system varies only marginally over the simulation time, and the same exponential initial condition is adopted in both cases. Consequently, $\int_{x_{\rm{min}}}^{x_{\rm{max}}}g(x,\tau)\, \mathrm{d}x\approx \int_0^{\infty}x\exp\,(-x)\, \mathrm{d}x=1$. We therefore directly analyze the error magnitudes in the following.

According to \cite{Liu2019}, the moments of the numerical solution are defined as,
\begin{equation}
    \begin{split}
M_{p,N}(\tau) &= \int_{x_{\min}}^{x_{\max}}x^{p-1}  g(x,\tau)\, \mathrm{d}x\\
&=\sum_{j=0}^{N-1}\int_{I_j}x^{p-1}g_j(x,\tau)\, \mathrm{d}x\\
&=\sum_{j=0}^{N-1}\sum_{i=0}^k g_j^i(\tau)\int_{I_j}x^{p-1}\phi_i(\xi_j(x))\, \mathrm{d}x.
    \end{split}
\end{equation}
Total mass of the system is the first moment $M_{1,N}(\tau) = \sum_{j=0}^{N-1} h_j g_j^0(\tau)$, which is exactly conserved by the scheme construction. The relatve errors on the moments are given by:
\begin{equation}
    e_{M_{p,N}}(\tau) = \frac{|M_{p,N}(\tau)-M_p(\tau)|}{M_p(\tau)}.
\end{equation}
where $M_p(\tau)$ is the moments of the analytic solution. To investigate the experimental order of convergence (EOC), we follow \cite{Liu2019},
\begin{equation}\label{eq:eoc}
    \mathrm{EOC} = \frac{\ln(\frac{e_N}{e_{2N}})}{\ln(2)},
\end{equation}
where $e_N$ is the error evaluated for $N$ mass bins and $e_{2N}$ for $2N$ mass bins.

\subsection{Pure Coagulation Test with Constant Kernel}\label{sec:pc}
\begin{figure*}
    \centering
    \includegraphics[width=1\linewidth]{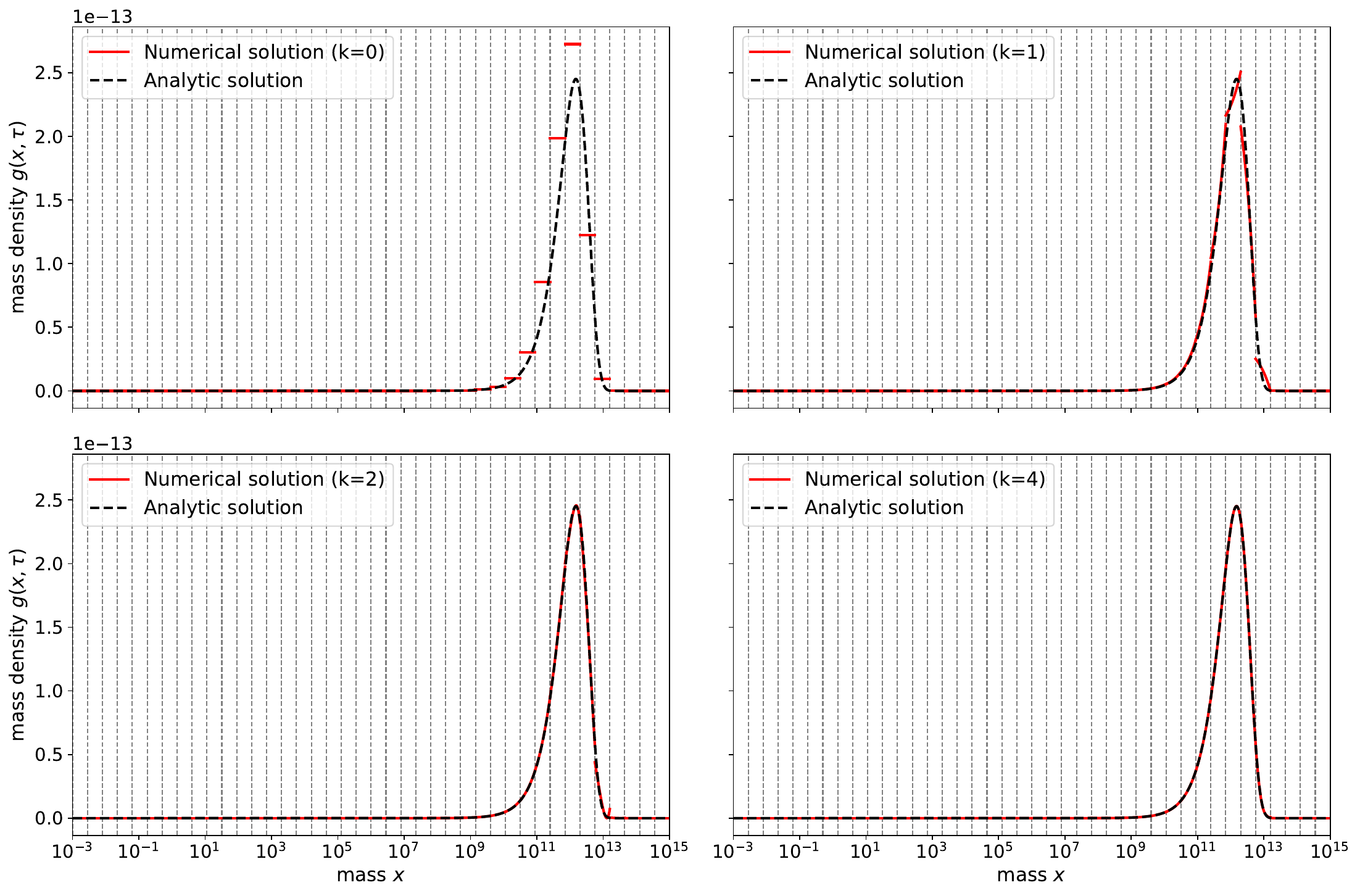}
    \caption{Numerical solutions for the pure coagulation test with constant kernel (Equation \ref{eq:fcoagsolution}) with $N=40$ cells and different $k$ at $\tau=3\times 10^{12}$. The Black dashed lines are the analytic solution for comparison; the vertical gray lines delimit the mass bins. The approximation improves for a higher polynomial degree.}
    \label{fig:coag}
\end{figure*}

Explicit analytic solutions are available for the coagulation-fragmentation equations with simple kernels and specific initial conditions, providing valuable benchmarks for validating numerical algorithms. For a constant coagulation kernel $\mathcal{K}_{\mathrm{coag}}(x,y)=1$ and an exponential initial condition of $f(x,0)=\exp\,(-x)$, the analytic solution of Equation \ref{eq:fcoag} is given by \citep{Scott1968},
\begin{equation}\label{eq:fcoagsolution}
    f(x,\tau)=\frac{4}{(2+\tau)^2}\exp\left( -\frac{2}{2+\tau}x\right).
\end{equation}
The mass density $g(x,\tau) = x f(x,\tau)$ reaches its peak at $x = 1 + \tau/2$, which increases over time due to dust growth. The prefactor $\frac{4}{(2+\tau)^2}$ reflects the decrease in number density as grains grow.

\begin{figure*}
    \includegraphics[width=1\textwidth]{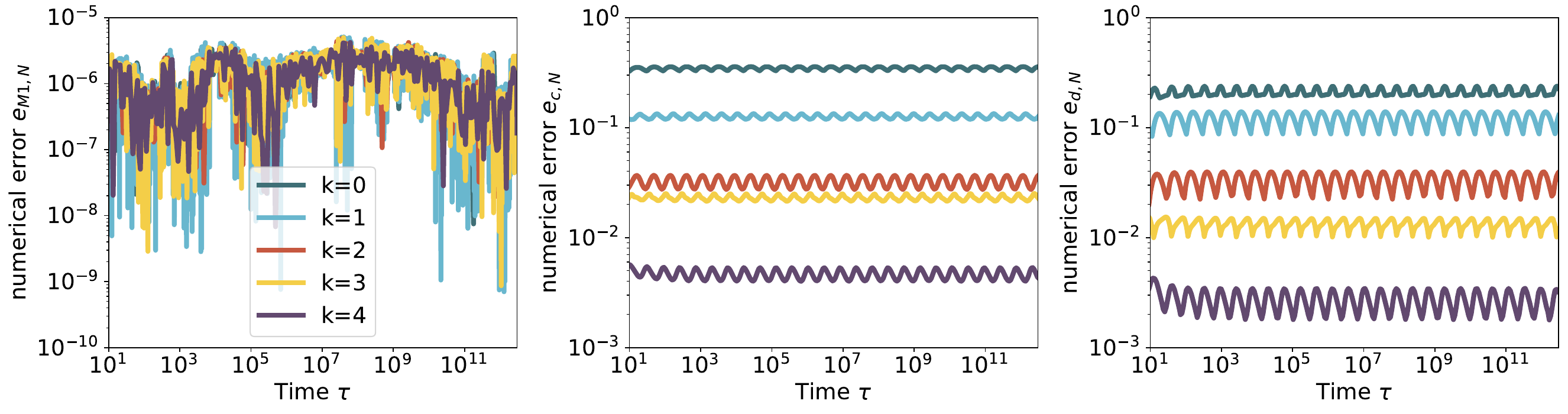}
    \caption{Time evolution of numerical errors for the pure coagulation test (consatnt kernel) with $N=40$ cells and different $k$. Errors remain bounded at large times (see Section \ref{sec:error} for error definitions).}
    \label{fig:coagerror}
\end{figure*}

Figure \ref{fig:coag} shows the numerical solutions at $\tau=3\times 10^{12}$ with $N=40$ mass bins and polynomial degree $k=0,1,2,4$. The computational domain is taken as $[10^{-3}, 10^{15}]$. For $k=0$, a uniform (piecewise constant) value is assigned within each mass bin, representing the cell-averaged mass density. In contrast, the higher-order scheme allows each mass bin to represent a polynomial distribution of degree up to $k$, enabling more accurate resolution of intra-bin variations.

Figure \ref{fig:coagerror} shows the evolution of numerical errors for the constant kernel with $N=40$ and $k$ ranging from $0$ to $4$. As shown in the left panel of Figure \ref{fig:coagerror}, absolute error on the total mass does not show any dependency on k and is demonstrated to be relatively small ($\sim 10^{-6}$). This holds true as long as the mass density peak does not exceed the upper boundary. We provide a more in-depth discussion regarding mass conservation in Section \ref{sec:mass conservation}. Both $e_{c,N}(\tau)$ and $e_{d,N}(\tau)$ decrease substantially with increasing $k$, an error of less than $1\%$ is already achieved at $k=4$.

In standard solvers, the gain and loss terms of Equation \ref{eq:fcoag} become huge and nearly equal when $x_{\mathrm{max}} \gtrsim 10^{14} x_{\mathrm{min}}$, causing near cancellation and severe loss of floating-point precision due to limited arithmetic precision. Consequently, traditional discretized solvers address this by rearranging the sums in the discrete Smoluchowski equation \citep{DD2005,Brauer2008,Stammler2022}. Our DG scheme, however, is based on the conservative equation and evolves the bin-averaged values rather than pointwise values, which inherently avoids such a cancellation issue. As demonstrated in this pure coagulation test, which covers 18 orders of magnitude in mass ($[10^{-3}, 10^{15}]$), Figure \ref{fig:coagerror} clearly shows that the numerical errors remain strictly bounded throughout the simulation.

\begin{figure}
    \includegraphics[width=\columnwidth]{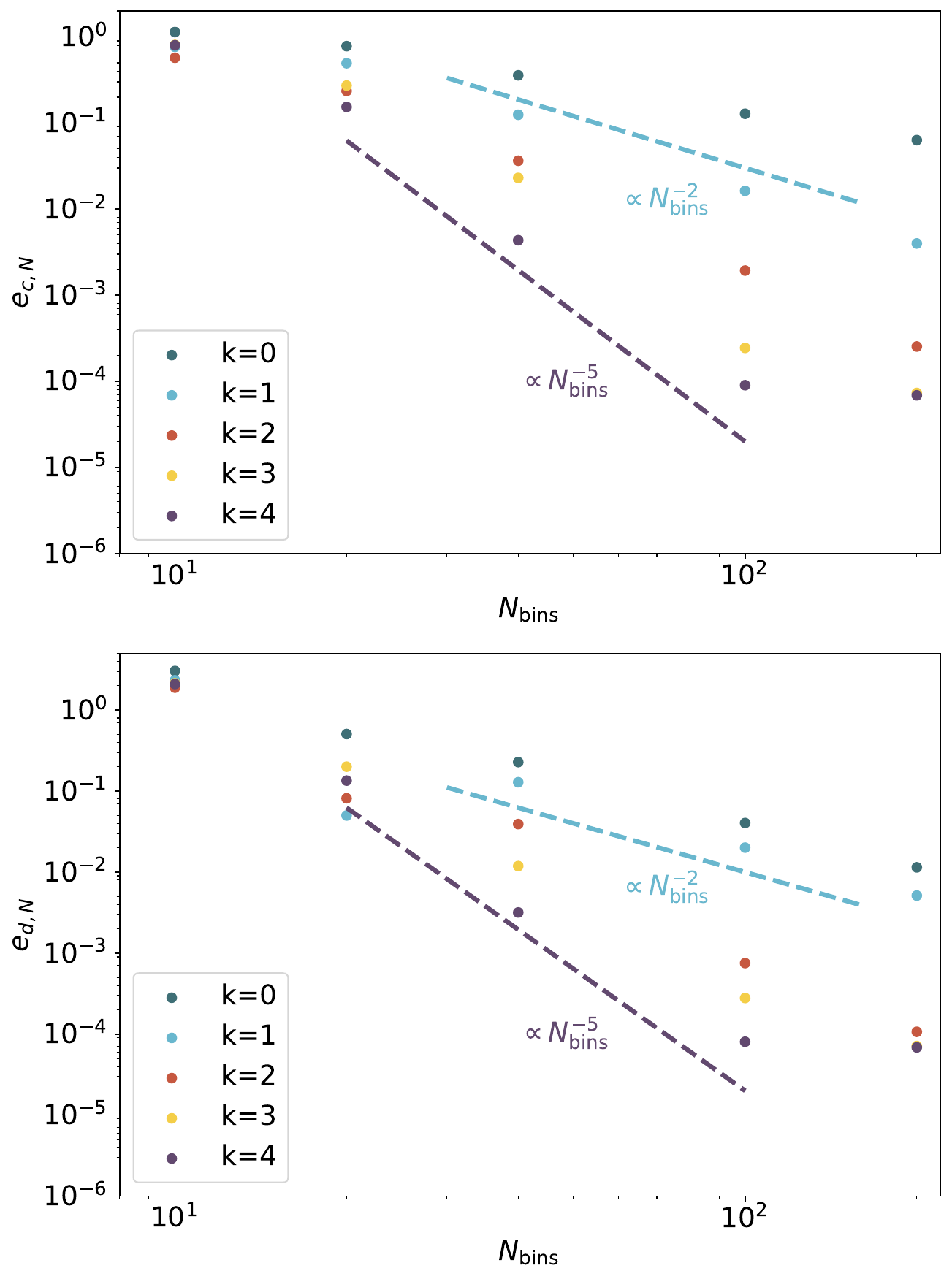}
    \caption{Numerical errors for $N=10, 20, 40, 100, 200$ and $k=0, 1, 2, 3, 4$ at $\tau=3\times 10^{12}$ for the pure coagulation test with constant kernel. The experimental order of convergence is $\mathrm{EOC}=k+1$.}
    \label{fig:coageoc}
\end{figure}

To examine the experimental order of convergence (EOC) of our numerical scheme, we perform pure coagulation simulations while systematically varying the number of mass bins $N$ and the polynomial degree $k$.Figure \ref{fig:coageoc} shows the numerical errors for $N=10, 20, 40, 100, 200$ and $k=0, 1, 2, 3, 4$ at $\tau=3\times 10^{12}$. The numerical scheme exhibits progressively better accuracy for larger mass bins $N$ and increased polynomial degree $k$, demonstrating an experimental order of convergence $\mathrm{EOC}=k+1$. This performance is consistent with the findings reported by previous studies. \citep{Liu2019,Lombart2021,Lombart2022,Lombart2026}

\subsection{Pure Coagulation Test with Additive Kernel}
In this subsection, we present a pure coagulation test with an additive kernel $\mathcal{K}_{\mathrm{coag}}(x,y)=x+y$ and compare our results with those obtained using the coagulation solver in {\tt Dustpy} \citep{Stammler2022}. Initially, only the first mass bin is populated, with a total number density $n_0=\int_0^{\infty}f(x,0)\, \mathrm{d}x$, while all other mass bins are set to zero. For the additive coagulation kernel, the solution of Equation \ref{eq:fcoag} is given by \citep{We1990},
\begin{equation}
\begin{split}
    f(x,\tau) &= \frac{n_0}{2x_{\rm{min}}\sqrt{\pi}(x/x_{\rm{min}})^{1.5}}\frac{T}{(1-T)^{0.75}} \\
    &\times \exp\left[-\frac{x}{x_{\rm{min}}}(1-\sqrt{1-T})^2\right],
\end{split}
\end{equation}
where $T = \exp(-n_0x_{\rm{min}}\tau)$.

\begin{figure}
    \centering
    \includegraphics[width=\columnwidth]{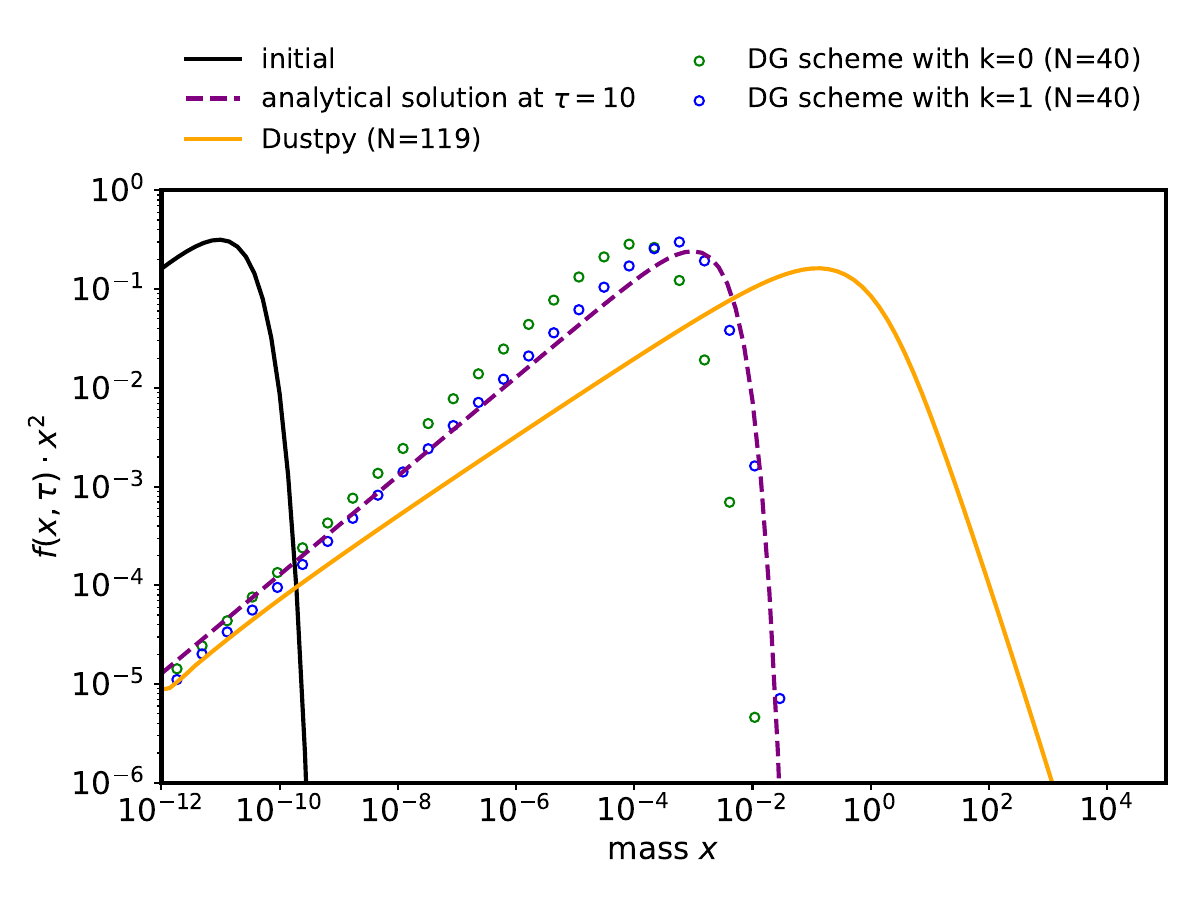}
    \caption{Comparison of coagulation solutions with an additive kernel. The orange line shows {\tt Dustpy} results, the purple dashed line is the analytical solution, and the green/blue dots are DG results with $N=40$ bins and $k=\{0,1\}$. }
    \label{fig:comparison}
\end{figure}

The results are shown in Figure \ref{fig:comparison}. The orange curve represents the simulation using the default mass resolution in {\tt Dustpy} (N=119), while the analytical solution is shown as a purple dashed line. Our discontinuous Galerkin (DG) simulations are performed with a smaller number of mass bins ($N=40$), using polynomial orders $k=0$ and $k=1$. For the DG results, we plot the cell-averaged values in each mass bin, shown as green and blue dots, respectively.

Despite the reduced mass resolution, the DG method accurately captures coagulation-driven growth and effectively suppresses over-diffusion, which is particularly severe for large mass intervals and mass-dependent kernels. The artificial grain growth visible in the {\tt Dustpy} results originates from its mass redistribution algorithm \citep{Stammler2022}: because the combined mass of two colliding particles in a sticking event generally does not coincide with a discrete mass grid point, the resulting mass must be distributed between neighboring bins. This procedure introduces artificial growth, as material is deposited into bins with masses larger than the true combined mass of the collision partners.

In summary, realistic coagulation kernels are highly sensitive to dust mass itself, and capturing dust growth from interstellar grains to pebbles requires a wide mass range. In conventional discretized methods like {\tt Dustpy}, this typically necessitates a few hundreds of mass bins to achieve accurate results. In contrast, the DG method mitigates the over-diffusion even with a coarse mass resolution and low polynomial order ($k=0$) \citep{Liu2019,Lombart2021}, making it a robust and efficient approach for modeling dust growth when the mass interval considered is large.

\subsection{Pure Fragmentation Test}\label{sec:purefrag}
\begin{figure*}
    \centering
    \includegraphics[width=1\linewidth]{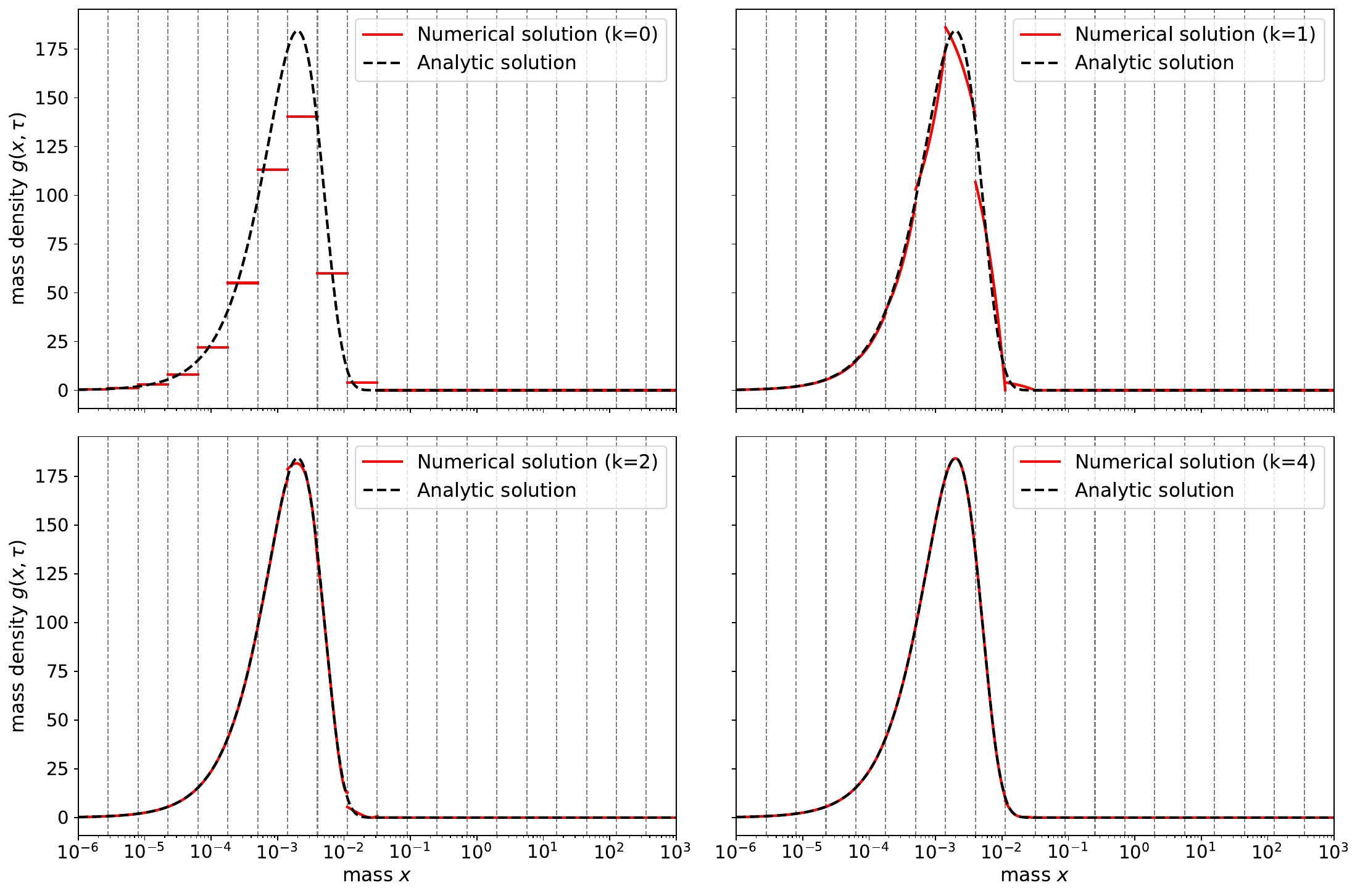}
    \caption{Numerical solutions for the pure fragmentation test (Equation \ref{eq:fragsolution}) with $N=20$ cells and different $k$ at $\tau=500$. The black dashed line is the analytic solution for comparison; the vertical gray lines delimit the bins. The approximation improves for a higher polynomial degree.}
    \label{fig:frag}
\end{figure*}

\begin{figure*}
    \includegraphics[width=1\textwidth]{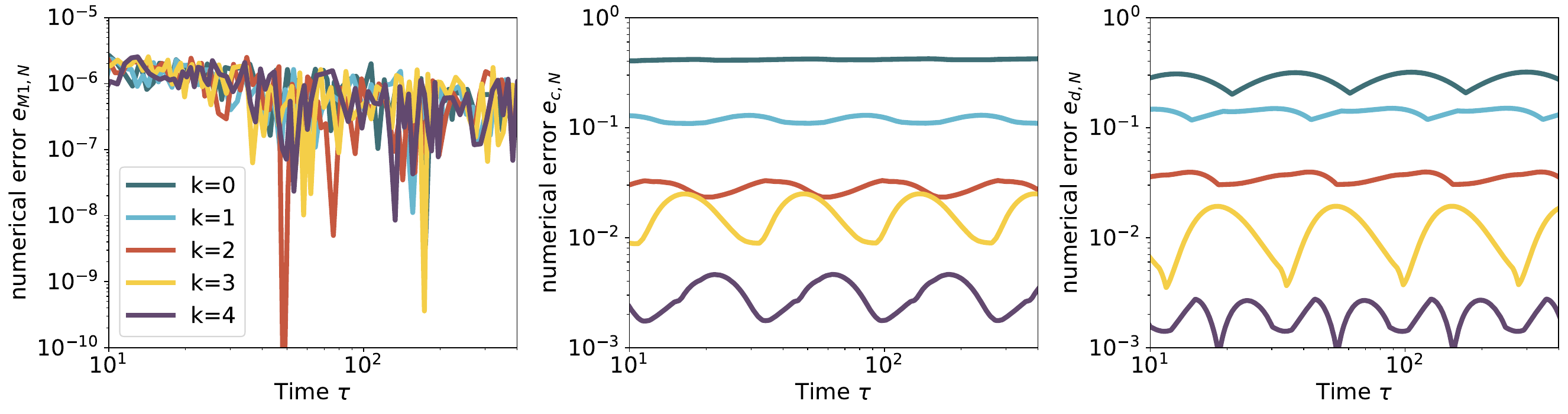}
    \caption{Time evolution of numerical errors for the pure fragmentation test with $N=20$ cells and different $k$. Errors remain bounded at large times. (see Section \ref{sec:error} for error definitions)}
    \label{fig:fragerror}
\end{figure*}
In the pure fragmentation test, we consider a symmetric breakage kernel,
\begin{equation}
    b(x,y,z) =\tilde b(x,y;z) + \tilde b(x,z;y),
\end{equation}
with which Equation \ref{eq:f24} reduces to Equation \ref{eq:f22}. The binary breakage takes a uniform fragments distribution,
\begin{equation}
    \tilde b(x,y;z) = \begin{cases}
    2/y & 0<x\le y, \\
    0 & x>y.
    \end{cases}
\end{equation}
The expected fragments number is $\int_0^y\tilde b(x,y;z)\,\mathrm{d}x = 2$, and the local mass conservation gives $\int_0^y x\tilde b(x,y;z)\,\mathrm{d}x=y$.
For a multiplicative kernel $\mathcal{K}_{\mathrm{frag}}(x,y)= xy$ and an exponential initial condition of $f(x,0)=\exp\,(-x)$, the analytic solution of Equation \ref{eq:f22} is given by \citep{Ziff1985},
\begin{equation}\label{eq:fragsolution}
    f(x,\tau)=(1+\tau)^2\exp\, \left(-x(1+\tau)\right).
\end{equation}
The mass density $g(x,\tau) = x f(x,\tau)$ peaks at $x = 1/(1+\tau)$, which decreases over time as a result of fragmentation processes. The prefactor $(1+\tau)^2$ captures the increase in number density as grains fragment into smaller pieces.

Figure \ref{fig:frag} shows the numerical solutions at $\tau=500$ for the multiplicative kernel with $N=20$ mass bins and polynomial degree $k=0,1,2,4$. The computational domain is taken as $[10^{-6}, 10^3]$. Figure \ref{fig:fragerror} shows evolutions of the numerical errors for the pure fragmentation simulation. Similar to the pure coagulation test, the numerical scheme achieves an error less than $1\%$ with $N = 20$ and $k = 4$. The experimental order of convergence is not presented here, as it closely resembles that of the pure coagulation case.

\subsection{Coupled Aggregation-Breakage}\label{sec:coupled}

While the benchmark tests presented above adopt highly idealized kernels (constant, additive, and multiplicative kernels), the collision rate in reality is considerably more complex. It depends not only on the sizes of the two colliding particles but also on their relative velocity, which plays a crucial role in determining both the collision frequency and the collisional outcome. 

In protoplanetary disks, the relative motion between dust grains can arise from various physical processes, including turbulent motion, Brownian motion, and systematic drift in the radial, vertical, or azimuthal directions, all of which are typically size-dependent \citep{Ormel2007,Birnstile2024}. To model the collisional evolution of dust aggregates, we consider three main collisional outcomes: coagulation, bouncing, and fragmentation. Instead of assuming a single relative velocity for all collisions, we allow the relative velocity to be drawn from a Maxwellian distribution with a size-dependent root-mean-square (rms) value.
\begin{equation}
    f_v = 4\pi v^2 \left(\frac{3}{2\pi \Delta v_{\mathrm{rms}}^2(x,y)}\right)^{3/2}\exp\left(-\frac{3v^2}{2 \Delta v_{\mathrm{rms}}^2(x,y)}\right).
\end{equation}

When the relative velocity exceeds the bouncing threshold $v_b$, particles bounce back rather than stick to each other. If the relative velocity further exceeds the fragmentation threshold $v_f$, particles undergo fragmentation. The probabilities of coagulation and fragmentation are respectively given by,
\begin{equation}
\begin{split}
    \beta_{\mathrm{coag}}(x,y)=\int_0^{v_{b}}f_v\,\mathrm{d}v = &\erf\left(\sqrt{\frac{3}{2\Delta v_{\mathrm{rms}}^2}}v_{b}\right)\\
    &-v_{b}\sqrt{\frac{6}{\pi \Delta v_{\mathrm{rms}}^2}}\exp\left( -\frac{3v_{b}^2}{2\Delta v_{\mathrm{rms}}^2} \right),
\end{split}
\end{equation}
\begin{equation}
\begin{split}
    \beta_{\mathrm{frag}}(x,y) = \int_{v_{f}}^{\infty}f_v\,\mathrm{d}v = &1-\erf\left(\sqrt{\frac{3}{2\Delta v_{\mathrm{rms}}^2}}v_{f}\right)\\
    &+v_{f}\sqrt{\frac{6}{\pi \Delta v_{\mathrm{rms}}^2}}\exp\left( -\frac{3v_{f}^2}{2\Delta v_{\mathrm{rms}}^2} \right),
\end{split}
\end{equation}
Then the coagulation and fragmentation kernels are denoted by,
\begin{equation}
    \mathcal K_{\mathrm{coag}}(x,y) =\sigma(x,y)\cdot \Delta v_{\mathrm{rms}}(x,y)\cdot \beta_{\mathrm{coag}}(x,y),
\end{equation}
\begin{equation}
    \mathcal K_{\mathrm{frag}}(x,y) =\sigma(x,y)\cdot \Delta v_{\mathrm{rms}}(x,y)\cdot \beta_{\mathrm{frag}}(x,y),
\end{equation}
where $\sigma(x,y)=\pi(x^{1/3}+y^{1/3})^2$ is the geometric cross section. $\beta_{\rm{coag}}(x,y) + \beta_{\rm{frag}}(x,y)$ may not be equal to 1 due to the bouncing effect.

\begin{figure}
    \centering
    \includegraphics[width=0.9\linewidth]{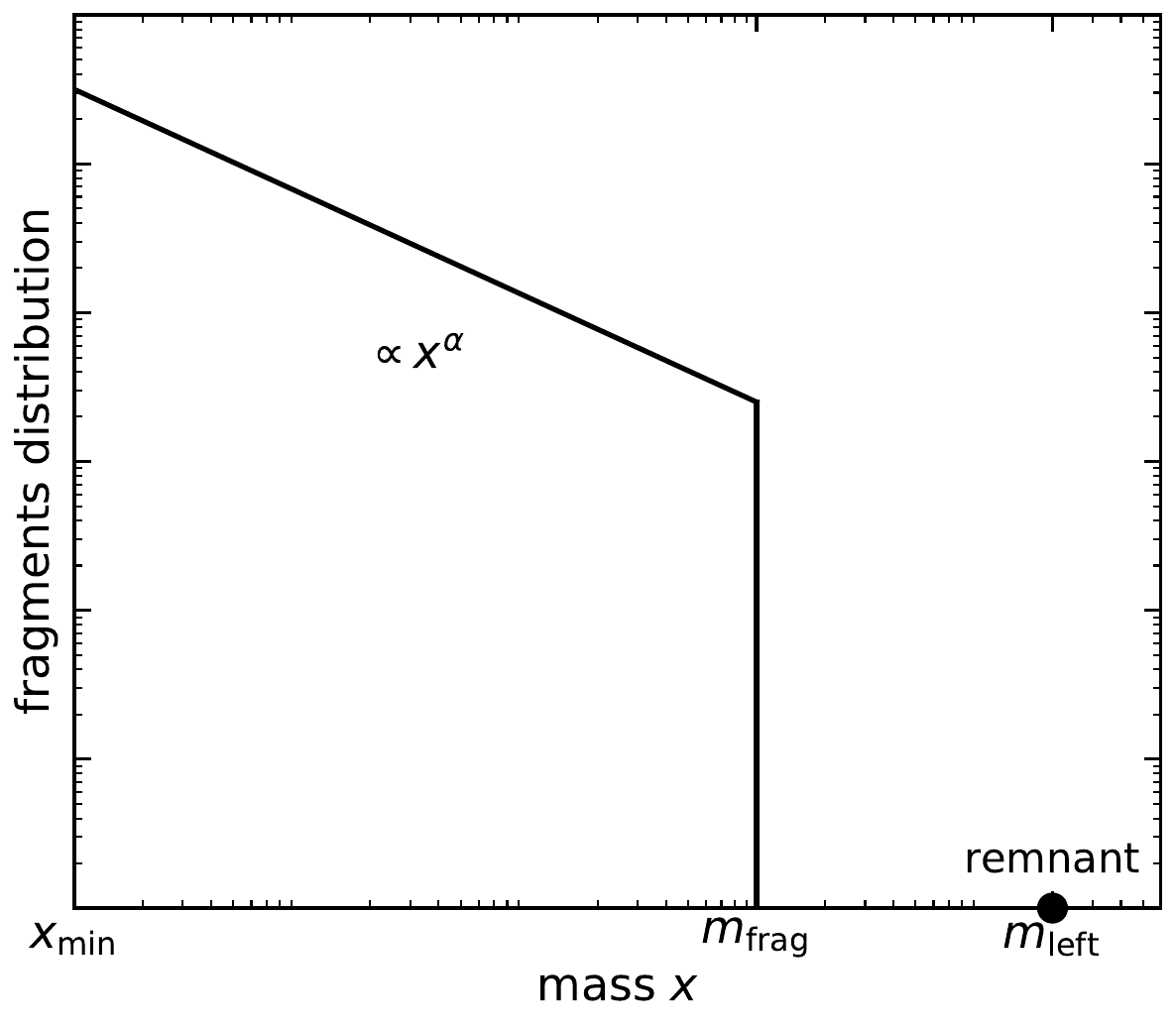}
    \caption{The mass distribution of fragmentation outcomes described by Equation \ref{eq:break}. The outcomes are divided into a continuous power-law fragment distribution ($x_{\min} \le x \le m_{\mathrm{frag}}$) and a discrete remnant particle ($m_{\mathrm{left}}$).}
    \label{fig:break}
\end{figure}

For fragmentation, we generally write the breakage kernel \citep{Kobayashi2010,Hirashita2021}:
\begin{equation}\label{eq:break}
    b(x,y,z) = Ax^{\alpha}+\delta(x-m_{\mathrm{left}}),
\end{equation}
where the first term represents a power law distribution of fragments with a total mass of $m_{\mathrm{frag}}$, the second term represents one remnant particle with mass of $m_{\mathrm{left}}=y+z-m_{\mathrm{frag}}$. Local mass conservation yields,
\begin{equation}
    \int_{x_{\min}}^{m_{\mathrm{frag}}}xAx^{\alpha}\, \mathrm{d}x= m_{\mathrm{frag}} \rightarrow A = \frac{(2+\alpha)m_{\mathrm{frag}}}{m_{\mathrm{frag}}^{2+\alpha}-x_{\min}^{2+\alpha}}.
\end{equation}
As illustrated in Figure \ref{fig:break}, we adopt a special case in our formulation where the upper cutoff of the power-law distribution (i.e., the maximum fragment mass) equals the total mass of all fragments, $m_{\mathrm{frag}}$ (for a general case, see Figure 1 of \cite{Kobayashi2010}). Several theoretical works obtained a power-law index $\alpha=-11/6\approx-1.83$ of the fragments distribution \citep{Dohnanyi1969,Jones1996}. The remnant mass should satisfy $m_{\mathrm{left}}\le y+z$, meaning that the collision between particles of mass $y$ and $z$ cannot create particles larger than their combined mass $y+z$. However, it is possible that a collision between particles of mass $y$ and $z$ might generate a fragment larger than $\max{(y,z)}$ due to the mass transfer from the smaller to the larger particle. 

The fraction of the total mass that ends up as fragments is typically modeled as a function of the mass ratio between the two colliding particles. Following \cite{Li2022}, we take into account the effects of cratering and mass transfer in addition to destructive fragmentation by,


\begin{equation}
\label{eq:mfrag}
m_{\mathrm{frag}} = (y+z) \cdot 
\begin{cases}
1 &  1 \le q \le 10 \quad \text{(destructive)} \\
\frac{1}{1+q} \cdot f_1(q) &  10 < q \le 12 \quad \text{(transition I)} \\
\frac{2}{1+q} &  12 < q \le 15 \quad \text{(cratering)} \\
\frac{1}{1+q} \cdot f_2(q) &  15 < q \le 50 \quad \text{(transition II)} \\
\frac{0.9}{1+q} &  q > 50 \quad \text{(mass transfer)}
\end{cases}
\end{equation}
where $q=\max{(y,z)}/\min{(y,z)}\ge1$ is the mass ratio. $f_1(q) = 6.5 + 4.5 \cos\left(\frac{q-10}{12-10}\pi\right)$ and $f_2(q)=1.45 + 0.55 \cos\left(\frac{q-15}{50-15}\pi\right)$ are taken to make the transition between different collisional outcomes smooth. A large mass ratio in collisions typically leads to the destruction of the smaller particle, while often lacking sufficient energy to fragment the larger one. In cases of extremely high mass ratios, part of the smaller particle's mass may even be deposited onto the larger one \citep{Wurm2005,Kothe2010,BukhariSyed2017}.

\begin{figure*}
    \centering
    \includegraphics[width=1\linewidth]{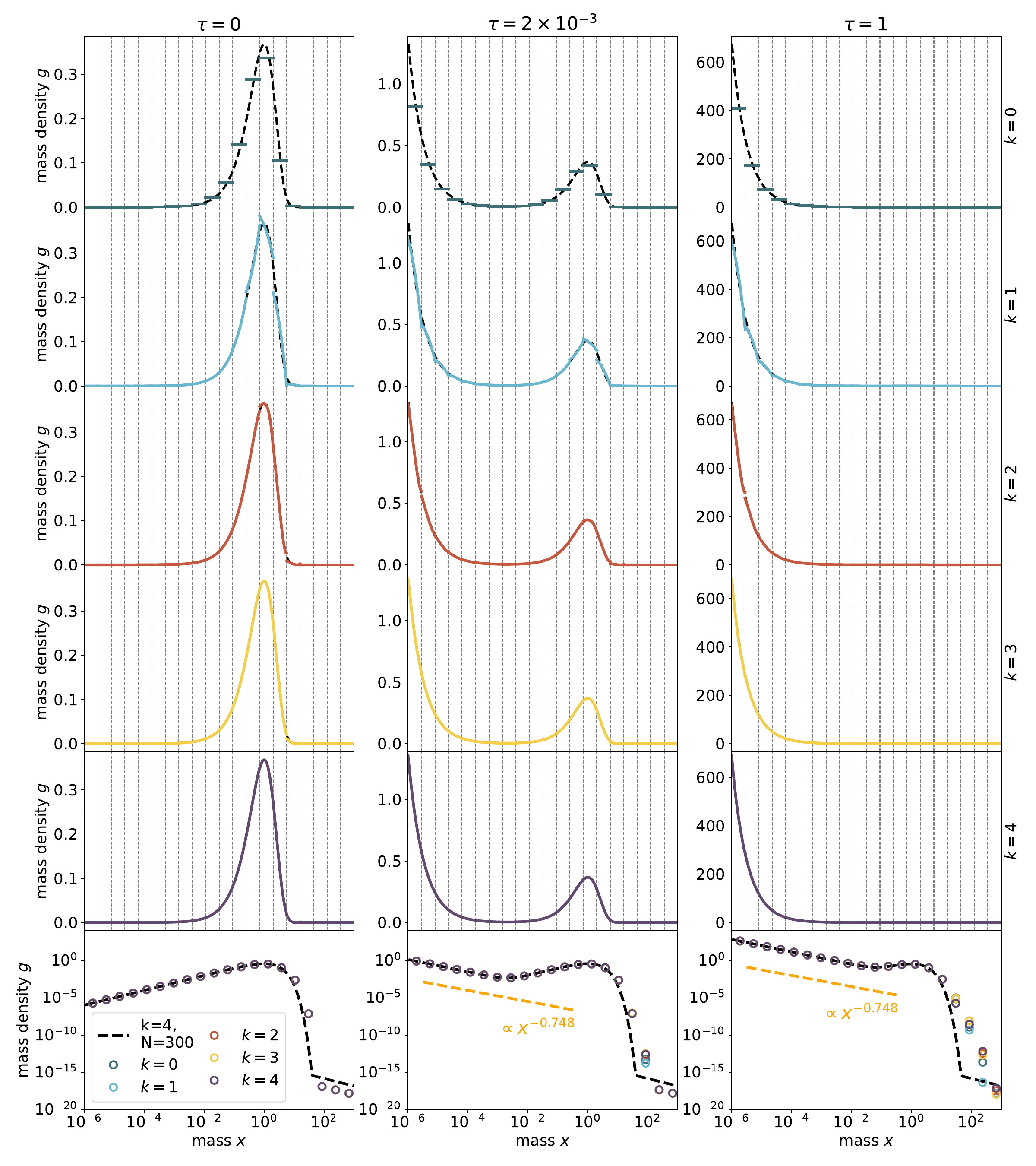}
    \caption{Numerical solutions for the case with coupled aggregation and breakage using $N=20$ mass bins. Vertical gray lines indicate the boundaries of each mass bin. Each row corresponds to a different polynomial degree $k$, while each column shows the solution at a different time $\tau$. The black dashed line represents the 
    reference solution computed with $N=300$ and $k=4$. The last row presents the same numerical solutions on a log-log scale, where the circles represent the averaged mass density for each bin.}
    \label{fig:cab}
\end{figure*}

To our knowledge, there's no known analytical solution for the general non-linear coagulation-fragmentation equation. Therefore, we present a toy model that incorporates the coupled aggregation and breakage processes, a Maxwellian distribution of relative velocities, and the general fragmentation prescription described by Equation \ref{eq:break} and \ref{eq:mfrag}, to show the capability of our code. For simplicity, we adopt a constant root-mean-square relative velocity of $\Delta v_{\mathrm{rms}}=0.5$, and fix the threshold velocities for bouncing and fragmentation at $v_b=0.1$, $v_f=1$, respectively. These values are chosen to illustrate the combined effects of coagulation and fragmentation.

We perform simulations using the same initial condition, $f(x,0)=\exp\,(-x)$ and adopt a power-law index of $\alpha=-1.83$. The computational domain is taken as $[10^{-6}, 10^3]$. Figure \ref{fig:cab} presents numerical results obtained using $N = 20$ mass bins, benchmarked against a reference solution computed with $N = 300$ and polynomial degree $k = 4$. This configuration exhibits negligible differences compared to solutions with even higher resolution.

Although increasing the polynomial order $k$ improves the accuracy of the intra-bin distribution within the smooth power-law regime, this internal refinement does not substantially enhance the accuracy of the mass distribution at the high-mass end. This limitation is fundamentally tied to the nature of polynomial basis functions. While polynomials excel at resolving smooth distributions, they struggle to capture the rapid, steep exponential drop-off at the high-mass end. As a consequence, beyond the truncation scale, higher-order approximations within individual bins cannot effectively compensate for the precision loss, making it difficult to further improve the accuracy of the flux computation. We defer to future work to address this issue.

In the bottom row of Figure \ref{fig:cab}, we observe that the mass density distribution at the lower-mass end scales as $g(x) \propto x^{-0.748}$. This result is consistent with the analytical scaling relation for a dust size distribution in coagulation-fragmentation equilibrium, $n(a) \propto a^q$, where $q = 2 - \frac{3}{2}(\nu - \alpha + 1)$ \citep{Birnstiel2011}. This translates to a mass distribution of $g(x) \propto x^\beta$, where $\beta = (q+4)/3 - 1$. For the power-law fragment distribution, we adopt an index of $\alpha = -1.83$, while the kernel index $\nu$ dictates how collision kernels scale with particle mass. For simplicity in this test case, we assume a constant root-mean-square relative velocity; thus, the mass dependence of the collision kernel arises solely from the geometric cross section, yielding $\nu = 2/3$, $q=-3.245$ and $\beta\approx-0.748$.


\section{Discussion}\label{sec:discussion}

\subsection{On the Choice of the Computational Domain} \label{sec:mass conservation}

In numerical simulations of dust coagulation and fragmentation, the computational domain is inherently restricted to a finite mass range $[x_{\mathrm{min}}, x_{\mathrm{max}}]$. The choice of these boundary limits can significantly influence the fidelity of the results. In this subsection, we discuss the physical and numerical considerations for selecting the lower and upper mass boundaries, highlighting how domain truncation impacts the dust size distribution.

In Section \ref{sec:coupled}, we present a test case featuring coupled coagulation and fragmentation processes, with a constant $\Delta v_{\mathrm{rms}}$. However, when accounting for the mass dependence of relative velocities, $\nu$ varies across physically distinct size regimes. This yields a piecewise power-law for the steady-state size distribution, with the index $q$ sequentially transitioning through $-2.5$, $-3.75$, and $-3.5$ as dust grain size increases \citep{Birnstiel2011}.
Furthermore, in drift-dominated environments where dust has not yet reached the fragmentation barrier, the distribution typically follows $q = -3$ \citep{Birnstile2024}. The mass density distribution scales as $a^{q+4}$, meaning large grains consistently dominate the total mass. Conversely, the dust surface area distribution, which is critical for electron adsorption and surface chemistry, scales as $a^{q+3}$, peaking at the transition between the Brownian motion and turbulent I regimes ($a_{\rm BT} \approx 0.1 - 1\ \mu\mathrm{m}$) \citep{Ormel2007}. 

As demonstrated, despite the inevitable domain truncation at the lower-mass end, our numerical scheme successfully preserves the correct steady-state power-law slopes within the resolved domain. In dust coagulation simulations, the monomer mass serves as a physically well-defined lower bound. Monomers are the fundamental constituent grains that assemble to form larger dust aggregates. They are typically assumed to be submicron-sized \citep{Birnstiel2011}. In practical computations, however, the appropriate choice of $x_{\min}$ remains problem-dependent. For simulations focusing primarily on dust hydrodynamics, resolving these submicron grains is unnecessary because they remain fully coupled to the gas. Instead, one can safely adopt a hydrodynamical lower limit. By defining a minimum Stokes number for gas coupling (the dimensionless product of the particle stopping time and the local Keplerian orbital frequency, e.g., $\mathrm{St}_{\min} \sim 10^{-2}$), the lower boundary $x_{\min}$ can be set to correspond to the particle size $a_{\rm{st}}$ below which the dust dynamically mimics the gas. Conversely, for simulations involving physics sensitive to the total dust surface area (e.g., non-ideal magnetohydrodynamics or surface chemistry), the computational domain must extend into the Brownian motion regime to fully capture the surface area peak, demanding a lower limit of $a \lesssim 0.1\ \mu\mathrm{m}$ for typical protoplanetary disks.



Having established the physical criteria for the lower size boundary, we next discuss the upper size limit from a purely numerical perspective, setting aside physical ceilings like the fragmentation barrier (which grains can surpass; see Section \ref{sec:hydro}). According to Equation \ref{eq:g0}, the change in mass within each mass bin $I_j$ results from the net effect of fluxes at $x_{j-1/2}$ and $x_{j+1/2}$. Under this construction, the total mass within the computational domain [$x_{\mathrm{min}}, x_{\mathrm{max}}$] is naturally conserved, provided that the fluxes at the boundaries are set to $0$. In Section \ref{sec:flux}, we have introduced two types of fluxes. The conservative flux with $\Delta x = -u+x_{\rm{min}}$ satisfies $F(x_{\min})=F(x_{\max})=0$, which helps conserve the total mass to machine precision.

A natural concern arises when the coagulation process dominates: since the mass flux at $x_{\max}$ is zero, mass tends to accumulate near the upper boundary as grains grow rather than forming even larger particles. In contrast, the non-conservative flux allows particles with masses exceeding $x_{\max}$ to form, however, the total mass within [$x_{\mathrm{min}}, x_{\mathrm{max}}$] is no longer conserved and instead decreases as grains grow. 

\begin{figure}
   \centering
   \includegraphics[width=1\linewidth]{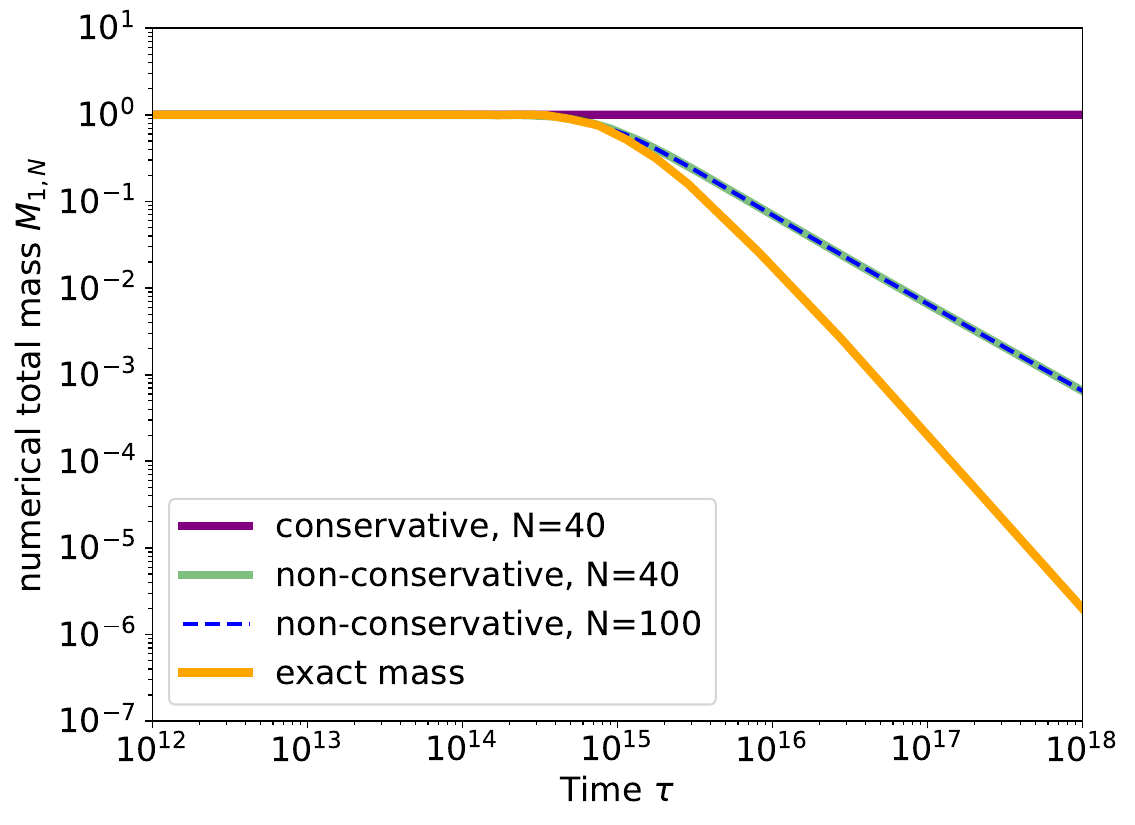}
   \caption{Total mass evolution for the pure coagulation test with $k=2$. Conservative and non-conservative flux results are shown in purple and green, respectively; the blue dashed line shows the higher-resolution non-conservative case. The orange line represents the evolution of $M_{1,N}$ from the exact solution.}
   \label{fig:mass}
\end{figure}

To illustrate the impact of the flux formulation, we simulate the pure coagulation test (see Section \ref{sec:pc} for the setup) over an extended time using both conservative and non-conservative fluxes with $N=40$ and $k=2$. In addition, a higher-resolution non-conservative case with $N=100$ is included. Figure \ref{fig:mass} shows the evolution of the total mass in each case, compared to the exact solution. The conservative-flux simulation (purple line) preserves the total mass $M_{1,N}$, as expected. In the non-conservative case, $M_{1,N}$ decreases over time due to particle growth beyond the upper mass boundary. Nonetheless, the results at both resolutions  ($N=40$ and $N=100$) still deviate from the exact solution.

This discrepancy is not caused by numerical error. Instead, it arises from the fact that we consider a finite computational domain [$x_{\mathrm{min}}, x_{\mathrm{max}}$], which excludes the grain mass distribution outside this range. However, the mass flux at a given point within the domain can, in principle, be influenced by the global distribution of $g(x,\tau)$ over $x\in(0, \infty)$. For example, a particle of mass $u<x$ may collide with another particle of mass $v>x_{\max}$, producing a particle of mass $u+v>x$, thereby contributing to the flux at $x$. This process is not accounted for---even when using the non-conservative flux---because $v>x_{\max}$ lies outside the domain of $g(x,\tau)$ considered in the simulation.

Consequently, special care must be taken in practical simulations when choosing the upper mass boundary. To ensure that the global solution is not artificially affected by boundary truncation, the upper mass boundary $x_{\max}$ should be placed sufficiently beyond the region where the dust mass density distribution is significant. In practice, this requires $x_{\max}$ to be large enough that the contribution from grains near and beyond the boundary is negligible over the timescale of interest.

\subsection{Toward Hydrodynamic Implementation}\label{sec:hydro}
Our understanding of protoplanetary disks has been advanced by increasingly sophisticated numerical models of gas (radiation- and magneto-) hydrodynamics. Recognizing the necessity of dust size evolution, recent simulations have begun to self-consistently incorporate dust dynamics and dust growth \citep[e.g.,][to name a few]{Li2020,Robinson2024,Kaufmann2025,Vaikundaraman2025,Lombart2026}.
To further advance this frontier, we introduce a novel coagulation-fragmentation solver that employs a high-order discontinuous Galerkin (DG) algorithm, capable of modeling a wide range of collisional outcomes as illustrated in Figure \ref{fig:model}.

\begin{figure}
    \centering
    \includegraphics[width=1\linewidth]{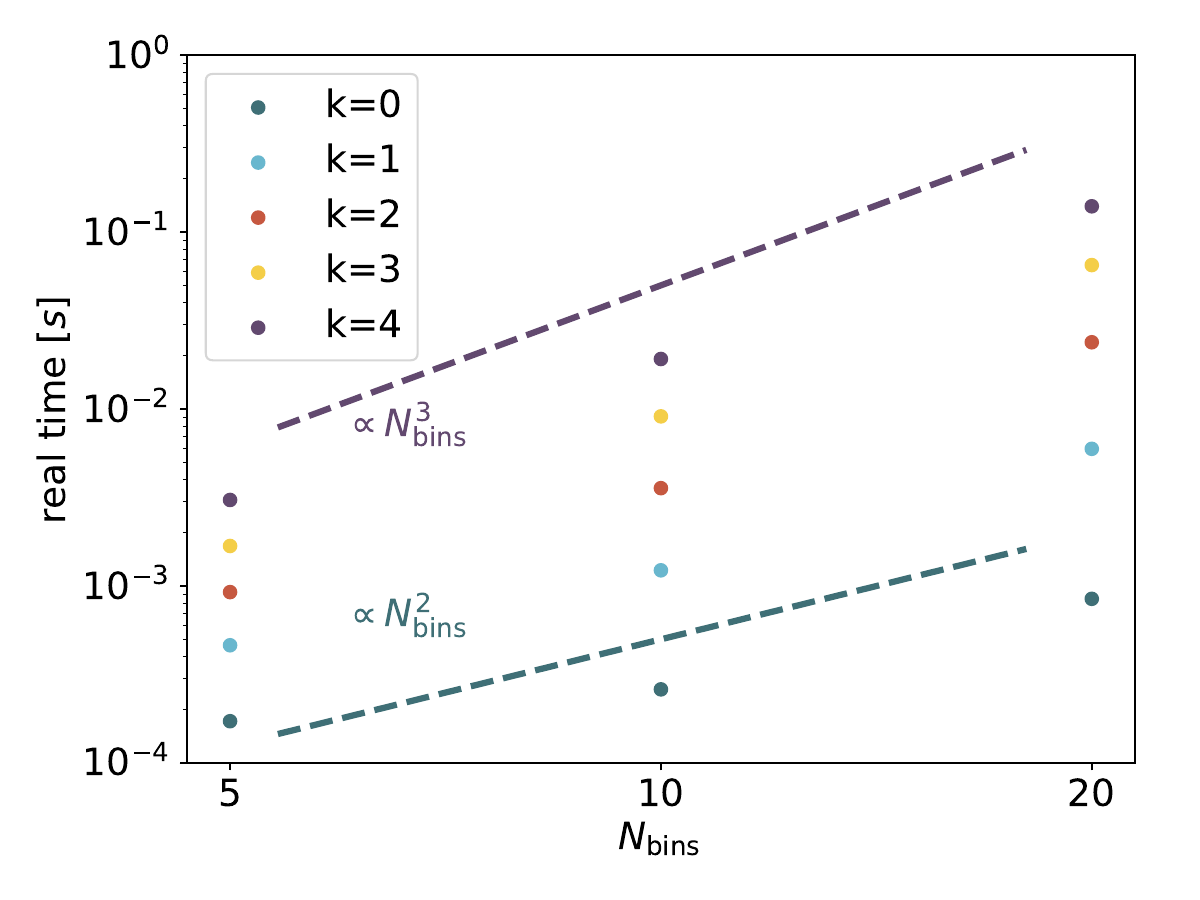}
    \caption{The averaged elapsed real time (in seconds) for the case with coupled aggregation and breakage. As expected from the algorithmic construction, the high-order method exhibits a scaling of $\propto N^3$, while the zeroth-order method which evolves only the cell-averaged mass density, shows a wall-clock time scaling of $\propto N^2$.}
    \label{fig:time}
\end{figure}

The averaged elapsed wall-clock time per time step of this DG solver is presented in Figure \ref{fig:time}. All simulations are performed for the case with coupled aggregation and breakage, as this represents a more realistic scenario when incorporated into hydrodynamical simulations. As before, the number of quadrature points chosen for a polynomial of degree $k$ is $Q=k+1$. The high-order DG method demonstrates a computational cost that scales as $\propto N^3$ with the number of mass bins, as anticipated based on the structure of the algorithm. Typically, it takes $\approx10^{-6}\, \mathrm{s}$ to update a cell in pure hydrodynamical simulations, which is significantly faster than the wall-clock time required for solving the coagulation-fragmentation equations---for example, $8.5\times 10^{-4}\, \mathrm{s}$ in the configuration of $N=20$ and $k=0$.

Nonetheless, we anticipate that incorporating dust collisional evolution into hydrodynamic codes, such as {\tt Athena++} and {\tt Guangqi} \citep{stone2020,huang2022,chen2026}, will not lead to a significant increase in computational cost. This is because, under realistic physical conditions in a typical protoplanetary disk, the timescale for dust growth is generally much longer than the hydrodynamic timescale. Simply consider a single-size approximation, the particle growth timescale is defined as \citep{Okuzumi2016},
\begin{equation}
    t_{\rm{growth}}=\frac{1}{\pi a_d^2n_d\Delta v_d},
\end{equation}
where $a_d$ denotes the representative dust size, $n_d$ and $\Delta v_d$ are the dust number density and collision velocity between dust grains. Taking dust-to-gas ratio to be $\Sigma_d / \Sigma_g=0.01$, $t_{\rm{growth}} \sim 10^4-10^5\, \rm{yr}$ at $100\, \rm{AU}$ and $\sim 1-100\, \rm{yr}$ at $1\, \rm{AU}$ \citep{Sato2016}, which is even longer than the corresponding orbital timescales. Therefore, the situation can be substantially alleviated by splitting the grain growth calculation from the evolution of hydrodynamics \citep{Ho2024}.

In addition to the computational efficiency, ensuring the physical fidelity of dust collisional evolution modeling requires particular attention to several key physical prescriptions. First, the relative velocity between dust grains must be accurately captured. In protoplanetary disks (PPDs), turbulent motion is the primary driver of relative velocity, which intrinsically scales with particle size \citep{Ormel2007,Birnstiel2011}. Equating this turbulent relative velocity to the fragmentation velocity results in an approximate maximum grain size \citep{Birnstiel2009},
\begin{equation}
    a_{\max} \approx \frac{2\Sigma_g}{\pi \alpha \rho_s}\cdot\frac{v_f^2}{c_s^2},
\end{equation}
where $\Sigma_g$ is the gas surface density, $\alpha$ is the turbulent viscosity parameter, $\rho_s$ is the bulk density of dust grains, and $c_s$ is the sound speed. 


Second, the inclusion of a bouncing barrier is essential for capturing early-stage growth limits. Previous studies have shown that bouncing can halt dust growth before the fragmentation process becomes significant. This prevents the continuous replenishment of small dust grains and results in a quasi-monodisperse size distribution peaking around $\sim 100\ \mu m$ \citep{Zsom2010,Dominik2024}. Such a specific size distribution is crucial for keeping PPDs observationally bright for millions of years \citep{Qian2025}.

Third, the precise formulation of the fragmentation kernel, specifically the treatment of mass transfer, can significantly alter growth outcomes. When mass transfer is allowed, larger grains ($a>a_{\rm{max}}$) can continue to grow even during collisions where the relative velocity exceeds $v_f$, In such scenarios, high-velocity impacts involving particles with a large mass ratio effectively contribute to grain growth beyond the standard fragmentation barrier, resembling a form of coagulation \citep{Windmark2012}.

In short, the complex interplay of these physical processes highlights the necessity of tracking the full size evolution of dust grains. While simplified model, such as {\tt Tripod} \citep{Pfeil2024,Kaufmann2025}, is highly efficient and successful in modeling global radial drift and dust growth, it fails to resolve the exact shape of size distributions when deviating from a single power law \citep{Eriksson2026}. With the method developed in this study, we can accommodate a wide range of collisional outcomes, and more realistic threshold conditions based on laboratory experiments and collision simulations can be readily implemented.


\subsection{Toward a Multidimensional Treatment}\label{sec:multidim}

Currently, GRACE-DG characterizes dust grains using a single parameter: mass, $x$. Astrophysical dust, however, exhibits a complex array of additional properties, including porosity, shape, chemical composition, and electric charge. These physical and chemical attributes directly govern collisional outcomes, aerodynamic coupling, and observational signatures. Capturing these effects requires extending the standard coagulation-fragmentation framework into a multidimensional population balance equation, which distributes the dust density across multiple internal coordinates.

Tracking distinct chemical species is vital for interpreting disk chemistry and meteoritic records. For instance, coupled models successfully explain water vapor depletion in upper disk layers \citep{Krijt2016}, observational signatures across CO snow lines \citep{Stammler2017}, and the isotopic dichotomy of refractory species like $^{54}$Cr between meteorite groups \citep{Homma2024}. To avoid a costly multidimensional grid, simulations often assume dust grains are macroscopic mixtures of various species. Numerically, the framework retains the 1D mass grid and solves a parallel set of coagulation-fragmentation equations to trace the partial mass of each constituent species.

Chemical composition can be efficiently tracked via parallel equations, and the surface composition of grains generally determines their fragmentation threshold (e.g., CO$_2$ ice, \citep{Musiolik2016a, Musiolik2016b}; H$_2$O ice, \citep{Musiolik2019}; silicates, \citep{Pillich2021}). However, incorporating porosity poses a highly non-trivial challenge because its physical effects are even more pervasive. Beyond influencing fragmentation threshold, porosity dictates the physical cross-sections and aerodynamic coupling of dust grains, thereby dynamically altering the collision kernels. Resolving this structural evolution fundamentally reshapes the macroscopic pathways of dust growth: highly porous aggregates present radically enlarged geometrical cross-sections, which accelerates coagulation and potentially enables them to safely bypass the radial drift barrier, particularly outside the snow line \citep{Ormel20072, Okuzumi20092, Okuzumi2012}. Conversely, subsequent collisional compaction and the onset of bouncing can abruptly stall these growth at millimeter scales \citep{Zsom2010, Zsom2011}.

Numerically managing this added dimensionality requires solving the generalized multidimensional Smoluchowski (coagulation) equation \citep{Kostoglou2001,Okuzumi20092,Watanabe2026}:
\begin{equation}
\begin{split}
\frac{\partial f(\tau, X)}{\partial \tau} = & \frac{1}{2} \int\int \mathrm{d}\, X_{\rm I} \mathrm{d}\, X_{\rm II} \mathcal{K}_{\mathrm{coag}}(X_{\rm I}, X_{\rm II}) \\
& \quad\quad\quad f(\tau, X_{\rm I}) f(\tau, X_{\rm II}) \delta(X - X_{\rm I+II}(X_{\rm I}, X_{\rm II})) \\
&- f(\tau, X) \int \mathrm{d}\, X_{\rm II} \mathcal{K}(X, X_{\rm II}) f(t, X_{\rm II}),
\end{split}
\end{equation}
which is a natural extension of Equation \ref{eq:fcoag}. The parameter vector $X = (X^{(1)}, X^{(2)}, \dots, X^{(d)})$ characterizes the dust aggregate across $d$ dimensions. The multidimensional Dirac delta function $\delta$ satisfies $\int \mathrm{d}X \delta(X - X_0) f(X) = f(X_0)$, while $X_{\rm I+II}(X_{\rm I}, X_{\rm II})$ is a function that returns the properties of the newly formed dust aggregate following the coagulation of two precursors.

Directly evaluating these multi-integrals on a fully discrete Eulerian mesh incurs a prohibitive computational cost that scales as $\mathcal{O}(N^{2d})$, where $N$ is the number of grid bins per dimension. To circumvent this severe bottleneck, several alternative approaches have been developed. Early works pioneered the use of Monte Carlo methods to simultaneously track mass and fractal dimension without relying on a rigid numerical grid \citep{Kostoglou2001,Ormel20072}. For Eulerian frameworks, computationally efficient moment equations are widely adopted. Tracking the zeroth- and first-order moments simultaneously evolves the grain size and volume distributions, directly yielding the aggregate porosity \citep{Okuzumi20092, Hirashita2021}. Most recently, algorithmic innovations have focused on directly accelerating the multidimensional integration itself; \citet{Watanabe2026} proposed a fast tree algorithm that groups distant nodes in phase space, successfully reducing the computational cost of multi-component coagulation from $\mathcal{O}(N^{2d})$ to $\mathcal{O}(d N^d \log N)$.

Extending the GRACE-DG to multidimensional spaces is mathematically natural. Taking a 2D mass-volume $(x,v)$ phase space as an example, within each two-dimensional cell $I_{j,m} = I_j \times I_m$, we can approximate the continuous distribution solution $g(x,v,\tau)$ by a linear combination of the tensor product of one-dimensional Legendre polynomials of degree at most $k$ in each dimension:
\begin{equation}
\begin{split}
\forall (x,v)\in I_{j,m} \quad & g(x,v,\tau) \approx g_{j,m}(x,v,\tau)\\
& = \sum_{p=0}^k \sum_{q=0}^k g_{j,m}^{p,q}(\tau)\phi_p(\xi_j(x))\phi_q(\eta_m(v)),
\end{split}
\end{equation}
where $g_{j,m}^{p,q}(\tau)$ is the component of $g_{j,m}(x,v,\tau)$ on the two-dimensional basis. Similarly, the local coordinates $\xi_j(x)$ and $\eta_m(v)$ map the physical domain $(x,v)\in I_{j,m}$ onto the reference square domain $(\xi,\eta) \in [-1,1] \times [-1,1]$.

Consequently, the calculation of the mass flux (e.g., Equation \ref{eq:Fcoag}) transitions from a double integral to a $2d$-dimensional integral. For example, it becomes a four-dimensional integral for the 2D mass-volume space. Although this direct integration fundamentally retains the steep $\mathcal{O}(N^{2d})$ computational scaling, the core strength of the DG method lies in its ability to aggressively shrink the base grid resolution, $N$. By leveraging high-order local approximations, the DG method substantially reduces the required grid resolution, $N$, effectively mitigating the multidimensional scaling penalty. Ultimately, benchmarking its computational efficiency and accuracy against state-of-the-art tree algorithm presents a compelling direction for future work.

\section{Conclusion}\label{sec:conclusion}
In this work, we developed a high-order DG method-based open-source code {\tt GRACE-DG} (GRAin Collisional Evolution - Discontinuous Galerkin) to solve the general non-linear coagulation-fragmentation equations. We derived a new mass flux formulation to accommodate general non-linear fragmentation processes (Equation \ref{eq:Ffrag}), such as the cratering and mass transfer (Figure \ref{fig:model}).

The DG scheme is formulated based on the conservative form of the Smoluchowski equations, which is positivity-preserving for both coagulation and fragmentation processes \citep{Liu2019,Lombart2021,Lombart2022,Lombart2024}, and can also alleviate the over-diffusion observed in existing discretized algorithms when solving the coagulation equation \citep{Stammler2022}.  We performed simulations combining aggregation and breakage effects (Section \ref{sec:coupled}), demonstrating good convergence properties and robustness. The incorporation of {\tt GRACE-DG} into hydrodynamic simulations is deferred to future studies.

\section*{Acknowledgments}
The authors are deeply grateful to the anonymous referee for the careful reading and thoughtful, constructive comments, which have substantially improved the quality and clarity of this manuscript. The authors thank Maxime Lombart, Ziyan Xu, Rixin Li, Pinghui Huang, Chris Ormel, Thomas Pfeil, and Linn Eriksson for constructive discussions. This work is supported by the National Science Foundation of China under grants No. 12233004, 12325304, 12342501, Tsinghua University Dushi Program, and Tsinghua University Initiative Scientific Research Program No. 20233080026. The Center of High performance computing at Tsinghua University provided the computational resources. We also acknowledge the Chinese Center for Advanced Science and Technology for hosting the Protoplanetary Disk and Planet Formation Summer School in 2024, which inspired this work.

\software{
    {\tt GRACE-DG} \citep{gracedg2026},
    {\tt DustPy} \citep{Stammler2022}
    }

\bibliography{dust}
\bibliographystyle{aasjournal}

\appendix
\section{conservative formulation}\label{sec:proof}
The general fragmentation of Equation \ref{eq:f24} can be equivalently written as,
\begin{equation}\label{eq:ap1}
\begin{split}
    \frac{\partial f(x,\tau)}{\partial \tau}=&\frac{1}{2}\int_x^{\infty}
    \mathrm{d}y\int_0^{y}\mathrm{d}z\,\mathcal{K}_{\mathrm{frag}}(y-z,z)b(x,y-z,z)f(y-z,\tau)f(z,\tau) -f(x,\tau)\int_0^{\infty}\mathcal K_{\mathrm{frag}}(x,y)f(y,\tau)\,\mathrm{d}y,
\end{split}
\end{equation}
We demonstrate its corresponding conservative form with respect to the mass density $g(x,\tau)=xf(x,\tau)$ is,
\begin{equation}\label{eq:ap2}
    \frac{\partial g(x,\tau)}{\partial \tau}+\frac{\partial F_{\mathrm{frag}}[g](x,\tau)}{\partial x}=0,
\end{equation}
\begin{equation}\label{eq:ap3}
\begin{split}
    F_{\mathrm{frag}}[g](x,\tau)&=\int_0^x\mathrm{d}u\int_{x-u}^{\infty}\mathrm{d}v\int_x^{u+v}\mathrm{d}w\, \frac{w}{v(u+v)} b(w,u,v)\mathcal K_{\mathrm{frag}}(u,v)g(u,\tau)g(v,\tau)\\
    &-\int_x^{\infty}\mathrm{d}u\int_0^{\infty}\mathrm{d}v\int_0^x\mathrm{d}w\, \frac{w}{v(u+v)} b(w,u,v)\mathcal K_{\mathrm{frag}}(u,v) g(u,\tau)g(v,\tau).
\end{split}
\end{equation}

Let's begin by applying the Leibniz integral rule to compute the partial derivative of $F_{\mathrm{frag}}[g](x,\tau)$ with respect to $x$,
\begin{equation}\label{eq:ap4}
\begin{split}
    \frac{\partial F_{\mathrm{frag}}[g](x,\tau)}{\partial x} &=\int_0^{\infty}\mathrm{d}v\int_{x}^{x+v}\mathrm{d}w\,\frac{w}{v(x+v)} b(w,x,v)\mathcal K_{\mathrm{frag}}(x,v)g(x,\tau)g(v,\tau)\\
    &-\int_0^x\mathrm{d}u\int_{x-u}^{\infty}\mathrm{d}v\, \frac{x}{v(u+v)} b(x,u,v)\mathcal K_{\mathrm{frag}}(u,v)g(u,\tau)g(v,\tau)\\
    &+\int_0^{\infty}\mathrm{d}v\int_{0}^{x}\mathrm{d}w\, \frac{w}{v(x+v)} b(w,x,v)\mathcal K_{\mathrm{frag}}(x,v)g(x,\tau)g(v,\tau)\\
    &-\int_x^{\infty}\mathrm{d}u\int_0^{\infty}\mathrm{d}v\, \frac{x}{v(u+v)} b(x,u,v)\mathcal K_{\mathrm{frag}}(u,v)g(u,\tau)g(v,\tau).
\end{split}
\end{equation}
The local mass conservation gives $\int_0^{x+v}wb(w,x,v)\, \mathrm{d}w=x+v$. By proceeding with a variable transformation and rearranging Equation \ref{eq:ap4}, we further obtain,
\begin{equation}
\begin{split}
    \frac{\partial F_{\mathrm{frag}}[g](x,\tau)}{\partial x} =xf(x,\tau)\int_0^{\infty}\mathcal K_{\mathrm{frag}}(x,y)f(y,\tau)\, \mathrm{d}y-x\int_{x}^{\infty}\mathrm{d}y\int_0^y \mathrm{d}z \frac{y-z}{y}\mathcal{K}_{\mathrm{frag}}(y-z,z)b(x,y-z,z)f(y-z,\tau)f(z,\tau).
\end{split}
\end{equation}
The symmetry with respect to the masses of the two colliding particles implies:
\begin{equation}
\begin{split}
    \int_{x}^{\infty}\mathrm{d}y\int_0^y \mathrm{d}z\, \frac{y-z}{y}\mathcal{K}_{\mathrm{frag}}(y-z,z)&b(x,y-z,z)f(y-z,\tau)f(z,\tau)=\int_{x}^{\infty}\mathrm{d}y\int_0^y \mathrm{d}z\, \frac{z}{y}\mathcal{K}_{\mathrm{frag}}(y-z,z)b(x,y-z,z)f(y-z,\tau)f(z,\tau)\\
    &=\frac{1}{2}\int_{x}^{\infty}\mathrm{d}y\int_0^y \mathrm{d}z\, \frac{y-z+z}{y}\mathcal{K}_{\mathrm{frag}}(y-z,z)b(x,y-z,z)f(y-z,\tau)f(z,\tau)\\
    &=\frac{1}{2}\int_{x}^{\infty}\mathrm{d}y\int_0^y \mathrm{d}z\, \mathcal{K}_{\mathrm{frag}}(y-z,z)b(x,y-z,z)f(y-z,\tau)f(z,\tau).
\end{split}
\end{equation}
Therefore, we demonstrate that Equation \ref{eq:ap2} is equivalent to,
\begin{equation}
    x\frac{\partial f(x,\tau)}{\partial \tau}+xf(x,\tau)\int_0^{\infty}\mathcal K_{\mathrm{frag}}(x,y)f(y,\tau)\, \mathrm{d}y-\frac{x}{2}\int_{x}^{\infty}\mathrm{d}y\int_0^y\mathrm{d}z\, \mathcal{K}_{\mathrm{frag}}(y-z,z)b(x,y-z,z)f(y-z,\tau)f(z,\tau)=0,
\end{equation}
which can be readily shown to match Equation \ref{eq:ap1}.

\section{fragmentation flux evaluation}\label{sec:gf}
Taking the breakage kernel in the form of Equation \ref{eq:break}, the conservative fragmentation flux of Equation \ref{eq:Ffragtrun} becomes,
\begin{equation}
    \begin{split}
        F_{\mathrm{frag}}^{\mathrm{tr}}[g](x,\tau)=&\int_{x_{\min}}^x\mathrm{d}u\int_{x-u+x_{\min}}^{x_{\max}-u+x_{\min}}\mathrm{d}v\int_{x}^{u+v}\mathrm{d}w\,\frac{w[Aw^{\alpha}+\delta(w-m_{\mathrm{left}})]}{v(u+v)} \mathcal K_{\mathrm{frag}}(u,v) g(u,\tau)g(v,\tau)\\
    -&\int_x^{x_{\max}}\mathrm{d}u\int_{x_{\min}}^{x_{\max}-u+x_{\min}}\mathrm{d}v\int_{x_{\min}}^{x}\mathrm{d}w\, \frac{w[Aw^{\alpha}+\delta(w-m_{\mathrm{left}})]}{v(u+v)}\mathcal K_{\mathrm{frag}}(u,v)g(u,\tau)g(v,\tau),
    \end{split}
\end{equation}
where $m_{\mathrm{left}}=m_{\mathrm{left}}(u,v)$ is the mass of the remnant, the domain of $Aw^{\alpha}$ is restricted to $w<m_{\mathrm{frag}}(u,v)$, which is given by Equation \ref{eq:mfrag}. Furthermore, we have $m_{\mathrm{frag}}+m_{\mathrm{left}}=u+v$, and $A=(2+\alpha)m_{\mathrm{frag}}/(m_{\mathrm{frag}}^{2+\alpha}-x_{\min}^{2+\alpha})$ derived from the local mass conservation.

The two outermost integrals over $u$ and $v$ are computed via Gaussian quadrature, following the flux evaluation approach presented in Section \ref{sec:flux}. Thus, the only remaining task is computing the innermost integral over $w$ for fixed $u=x_l^\alpha$ and $v=x_m^\beta$:
\begin{equation}
    \int_{x}^{u+v} w [Aw^{\alpha}+\delta(w-m_{\mathrm{left}})]\, \mathrm{d}w = \begin{cases}
    \frac{A}{2+\alpha}\cdot (m_{\mathrm{frag}}^{\alpha+2}-x^{\alpha+2}) + m_{\mathrm{left}} & x<m_{\mathrm{frag}},\quad  x<m_{\mathrm{left}}. \\
    \frac{A}{2+\alpha}\cdot (m_{\mathrm{frag}}^{\alpha+2}-x^{\alpha+2}) & x<m_{\mathrm{frag}},\quad  x>m_{\mathrm{left}}. \\
    m_{\mathrm{left}} & x>m_{\mathrm{frag}},\quad  x<m_{\mathrm{left}}. \\
    0 & x>m_{\mathrm{frag}},\quad  x>m_{\mathrm{left}}.
    \end{cases}
    \label{eq:inte1}
\end{equation}
\begin{equation}
    \int_{x_{\min}}^{x} w [Aw^{\alpha}+\delta(w-m_{\mathrm{left}})]\, \mathrm{d}w =\begin{cases}
    \frac{A}{2+\alpha}\cdot (x^{\alpha+2}-x_{\min}^{\alpha+2})  & x<m_{\mathrm{frag}},\quad  x<m_{\mathrm{left}}. \\
    \frac{A}{2+\alpha}\cdot (x^{\alpha+2}-x_{\min}^{\alpha+2}) + m_{\mathrm{left}} & x<m_{\mathrm{frag}},\quad  x>m_{\mathrm{left}}. \\
    \frac{A}{2+\alpha}\cdot (m_{\mathrm{frag}}^{\alpha+2}-x_{\min}^{\alpha+2}) & x>m_{\mathrm{frag}},\quad  x<m_{\mathrm{left}}. \\
    \frac{A}{2+\alpha}\cdot (m_{\mathrm{frag}}^{\alpha+2}-x_{\min}^{\alpha+2}) + m_{\mathrm{left}} & x>m_{\mathrm{frag}},\quad  x>m_{\mathrm{left}}.
    \label{eq:inte2}
\end{cases}
\end{equation}

Then, the integral of mass flux over $I_j$,
\begin{equation}
    \begin{split}
        \int_{I_j}F_{\mathrm{frag}}^{\mathrm{tr}}[g](x,\tau)\partial_x\phi \, \mathrm{d}x=&\int_{I_j}\mathrm{d}x\int_{x_{\min}}^x\mathrm{d}u\int_{x-u+x_{\min}}^{x_{\max}-u+x_{\min}}\mathrm{d}v\int_{x}^{u+v}\mathrm{d}w\, \frac{w[Aw^{\alpha}+\delta(w-m_{\mathrm{left}})]}{v(u+v)} K_{\mathrm{frag}}(u,v)g(u,\tau)g(v,\tau)\partial_x\phi \\
        -&\int_{I_j}\mathrm{d}x\int_x^{x_{\max}}\mathrm{d}u\int_{x_{\min}}^{x_{\max}-u+x_{\min}}\mathrm{d}v\int_{x_{\min}}^x\mathrm{d}w\,  \frac{w[Aw^{\alpha}+\delta(w-m_{\mathrm{left}})]}{v(u+v)} \mathcal K_{\mathrm{frag}}(u,v)g(u,\tau)g(v,\tau)\partial_x\phi
    \end{split}
\end{equation}

By reordering the integration sequence, we get to,
\begin{equation}
    \begin{split}
        \int_{I_j}F_{\mathrm{frag}}^{\mathrm{tr}}[g](x,\tau)\partial_x\phi \, \mathrm{d}x=&\int_{x_{\min}}^{x_{\max}}\mathrm{d}u\int_{x_{\min}}^{x_{\max}-u+x_{\min}}\mathrm{d}v\int_{I_j\cap [u, m_{\mathrm{frag}}]}\mathrm{d}x\,\frac{A (m_{\mathrm{frag}}^{\alpha+2}-x^{\alpha+2})}{2+\alpha} \partial_x\phi \cdot \frac{\mathcal K_{\mathrm{frag}}(u,v)g(u,\tau)g(v,\tau)}{v(u+v)}\\
        -&\int_{x_{\min}}^{x_{\max}}\mathrm{d}u\int_{x_{\min}}^{x_{\max}}\mathrm{d}v\int_{I_j\cap [x_{\min},u]}\mathrm{d}x\,\frac{A (x^{\alpha+2}-x_{\min}^{\alpha+2})}{2+\alpha}\partial_x\phi \cdot \frac{\mathcal K_{\mathrm{frag}}(u,v)g(u,\tau)g(v,\tau)}{v(u+v)}\\
        +&\int_{x_{\min}}^{x_{\max}}\mathrm{d}u\int_{x_{\min}}^{x_{\max}}\mathrm{d}v\int_{I_j\cap [m_{\mathrm{frag}},u]}\mathrm{d}x\,\frac{A (x^{\alpha+2}-m_{\mathrm{frag}}^{\alpha+2})}{2+\alpha}\partial_x\phi \cdot \frac{\mathcal K_{\mathrm{frag}}
        (u,v)g(u,\tau)g(v,\tau)}{v(u+v)}\\
        +&\int_{x_{\min}}^{x_{\max}}\mathrm{d}u\int_{x_{\min}}^{x_{\max}-u+x_{\min}}\mathrm{d}v\int_{I_j\cap [u, m_{\mathrm{left}}]}\mathrm{d}x\, m_{\mathrm{left}}\partial_x\phi \cdot \frac{\mathcal K_{\mathrm{frag}}(u,v)g(u,\tau)g(v,\tau)}{v(u+v)}\\
        -&\int_{x_{\min}}^{x_{\max}}\mathrm{d}u\int_{x_{\min}}^{x_{\max}}\mathrm{d}v\int_{I_j\cap [m_{\mathrm{left}},u]}\mathrm{d}x\, m_{\mathrm{left}}\partial_x\phi \cdot \frac{\mathcal K_{\mathrm{frag}}(u,v)g(u,\tau)g(v,\tau)}{v(u+v)}.
    \end{split}
\end{equation}

With all conditional constraints on the integral bounds for $x$ established, we can now similarly apply Gaussian quadrature for evaluating $\int_{I_j}F_{\mathrm{frag}}^{\mathrm{tr}}[g](x,\tau)\partial_x\phi\, \mathrm{d}x$.

\end{CJK*}
\end{document}